\documentclass[prd,preprintnumbers,floatfix,
nofootinbib,superscriptaddress]{revtex4}

\usepackage{float}
\usepackage{nicefrac}
\usepackage{mathtools}
\usepackage{amsfonts} 
\usepackage{amssymb} 
\usepackage{amsmath} 
\usepackage{bbold}
\usepackage{bm} 
\usepackage{graphicx} 
\usepackage{subfigure} 
\usepackage{array} 
\usepackage{dcolumn} 
\usepackage{bm} 
\usepackage{latexsym} 
\usepackage{longtable} 
\usepackage{hyperref} 
\usepackage{verbatim}
\usepackage{epsfig}
\usepackage{slashed}
\usepackage{color}

\usepackage{glossaries-extra}
\setabbreviationstyle[acronym]{long-short}
\glssetcategoryattribute{acronym}{nohyperfirst}{true}
\renewcommand{\vec}[1]{\mathbf{#1}}
\newcommand\bra[1]{\langle #1 \vert}
\newcommand\ket[1]{\vert #1 \rangle}

\newcommand\braket[2]{\langle #1 \vert #2 \rangle}

\newcommand{\matS}[0]{\mathbb S}

\newcommand{\cA}[0]{\mathcal A}
\newcommand{\cB}[0]{\mathcal B}

\newcommand{\cD}[0]{\mathcal D}

\newcommand{\cI}[0]{\mathcal I}
\newcommand{\cK}[0]{\mathcal K}

\newcommand{\cM}[0]{\mathcal M}
\newcommand{\cO}[0]{\mathcal O}

\newcommand{\cS}[0]{\mathcal S}

\newcommand{\cY}[0]{\mathcal Y}

\newcommand{\off}[0]{{\rm off}}

\newcommand{\wh}[0]{\widehat}
\newcommand{\wt}[0]{\widetilde}
\newcommand{\mc}[0]{\mathcal}
\newcommand{\df}[0]{\mathrm{df}}

\newcommand{\uu}[0]{{(u,u)}}

\newcommand{\Kdf}[0]{{\cK_{\df,3}}}

\newcommand{\PV}[0]{{\mathrm{PV}}}

\newcommand{\ID}[0]{{\mathrm{ID}}}
\newcommand{\nd}[0]{{\mathrm{nd}}}

\newcommand{\HSQCa}[0]{Hansen:2014eka}
\newcommand{\HSQCb}[0]{Hansen:2015zga}

\newcommand{\BHSQC}[0]{Briceno:2017tce}
\newcommand{\BHSnum}[0]{Briceno:2018mlh}
\newcommand{\BHSK}[0]{Briceno:2018aml}
\newcommand{\dwave}[0]{Blanton:2019igq}
\newcommand{\largera}[0]{Romero-Lopez:2019qrt}

\newcommand{\RtoK}[0]{Jackura:2019bmu}

\newcommand{\isospin}[0]{Hansen:2020zhy}

\newcommand{\BSQC}[0]{Blanton:2020jnm}
\newcommand{\BSequiv}[0]{Blanton:2020gha}
\newcommand{\BSnondegen}[0]{Blanton:2020gmf}

\newcommand{\Akakia}[0]{Hammer:2017uqm}
\newcommand{\Akakib}[0]{Hammer:2017kms}

\newcommand{\DDK}[0]{Pang:2020pkl}

\newcommand{\MDpi}[0]{Mai:2018djl}

\newcommand{\MD}[0]{Mai:2017bge}
\newcommand{\MDHH}[0]{Mai:2019fba}

\newcommand{\HSrev}[0]{Hansen:2019nir}
\newcommand{\Akakirev}[0]{Rusetsky:2019gyk}
\newcommand{\MDRrev}[0]{Mai:2021lwb}
\newcommand{\Maiisobar}[0]{Mai:2017vot}
\newcommand{\isobar}[0]{Jackura:2018xnx}

\newcommand{\ThreeQCDNumerics}[0]{%
Mai:2018djl,
Horz:2019rrn,
Blanton:2019vdk,
Culver:2019vvu,
Mai:2019fba,
Fischer:2020jzp,
Hansen:2020otl,
Alexandru:2020xqf,
Brett:2021wyd}

\newcommand{\ThreeBody}[0]{%
Briceno:2012rv,
Polejaeva:2012ut,
Hansen:2014eka,
Hansen:2015zga,
Briceno:2017tce,
Hammer:2017uqm,
Hammer:2017kms,
Mai:2017bge,
Briceno:2018aml,
Briceno:2018mlh,
Pang:2019dfe,
Jackura:2019bmu,
Blanton:2019igq,
Briceno:2019muc,
Romero-Lopez:2019qrt,
Pang:2020pkl,
Blanton:2020gha,
Blanton:2020jnm,
Romero-Lopez:2020rdq,
Blanton:2020gmf,
Muller:2020vtt}

\newacronym{CMF}{CMF}{center-of-momentum frame}

\begin{document}


\title{Three-particle finite-volume formalism for $\pi^+\pi^+ K^+$ and related systems}

\author{Tyler D. Blanton}
\email[e-mail: ]{blanton1@uw.edu}
\affiliation{Physics Department, University of Washington, Seattle, WA 98195-1560, USA}

\author{Stephen R. Sharpe}
\email[e-mail: ]{srsharpe@uw.edu}
\affiliation{Physics Department, University of Washington, Seattle, WA 98195-1560, USA}


\date{\today}

\begin{abstract}
We consider three-particle systems consisting of
two identical particles and a third that is different, with all being spinless.
Examples include $\pi^+\pi^+ K^+$ and $K^+K^+\pi^+$.
We derive the formalism necessary to extract 
two- and three-particle infinite-volume scattering amplitudes
from the spectrum of such systems in finite volume.
We use a relativistic formalism based on an all-orders diagrammatic analysis
in generic effective field theory,
adopting the methodology used recently to study the case of three nondegenerate particles.
We present both a direct derivation, and also a cross-check based on an appropriate limit 
and projection of the fully nondegenerate formalism.
We also work out the threshold expansions for the three-particle K matrix that will be
needed in practical applications, both for systems with two identical particles plus a third,
and also for the fully nondegenerate theory.

 \end{abstract}


\nopagebreak

\maketitle


\section{Introduction\label{sec:intro}}

Recent advances in theoretical formalism~\cite{\ThreeBody}
and its application to the results of numerical simulations~\cite{\ThreeQCDNumerics}
have allowed the first studies of three-particle interactions in lattice QCD.
Many three-particle systems of interest cannot, however,
be treated using existing formalism, either because
the particles have spin (e.g.~systems including nucleons),
because multiple three-particle channels are involved (e.g.~$\pi^+\pi^0 K^+$ and $\pi^+ \pi^+ K^0$),
or because two of the three particles are identical and the kinematics is relativistic
(examples given below).
In this paper we extend the formalism to the latter systems (with all particles spinless),
which hereinafter we refer to as ``$2+1$'' systems.

Three main approaches have been followed in the derivation of three-particle finite-volume formalism.
That adopted here is based on an all-orders diagrammatic analysis in a generic relativistic
effective field theory (EFT). This is denoted the RFT approach, and has been developed in 
Refs.~\cite{\HSQCa,\HSQCb,\BHSQC,\BHSK,\largera,\isospin,\BSQC,\BSnondegen}.
The alternative approaches are based on nonrelativistic effective field theory
(NREFT)~\cite{\Akakia,\Akakib,\DDK}
and on a finite-volume extension of unitary (FVU approach)~\cite{\MDpi,\MD,\MDHH}.
To date, only the RFT approach has been developed for general relative angular momenta.
The equivalence of the RFT and FVU approaches
for systems of three identical scalars is shown in Ref.~\cite{\BSequiv}.
For an overview of the present status see the reviews in Refs.~\cite{\HSrev,\Akakirev,\MDRrev}.

The formalism that has been derived to date applies only for spinless particles, 
and assumes that the particles are either 
identical~\cite{\HSQCa,\HSQCb,\Akakia,\Akakib,\MDpi,\MD,\BSQC},
degenerate and related by isospin symmetry~\cite{\isospin},
or distinguishable and, in general, nondegenerate~\cite{\BSnondegen}.
This allows applications to 
extremal charge configurations of three particles ($\pi^+\pi^+\pi^+$, $K^-K^-K^-$, etc.),
to general three-pion systems in the isospin-symmetric limit,
and to a limited number of nondegenerate systems 
(e.g.~$D_s^+ D^0 \pi^-$ and $D_s^+ D^0 \pi^-$).
The only exception to the above-described limitations is the recent extension of the NREFT formalism,
in the s-wave limit, to the $DDK$ system~\cite{\DDK}.

The extension presented here allows the application of the formalism
to several systems of phenomenological interest in QCD, including
$\pi^+\pi^+ K^+$, $\pi^+\pi^+ \overline K^0$, $\pi^+K^+K^+$, and $\pi^+\pi^+ D_s^\pm$.
The key property of these examples is that quark (or antiquark) exchange between the particles
leads either to the same channel (as in the first three examples) or to a heavier channel
(e.g.~$\pi^+\pi^+ D_s \to \pi^+ K^+ D^+$).
This allows one to consider only a single three-particle channel for an appropriate kinematic range.
We also note that lattice QCD calculations can determine the finite-volume spectra 
for these examples with only minor extensions of the setups used for the studies done to date,
which involve $3\pi^+$ or $3 K^-$ (or the charge-conjugate systems).
Thus we hope that the formalism developed here will be of immediate practical value.

We derive the results presented here using the simplified method presented 
in Ref.~\cite{\BSnondegen}, itself based on an approach developed in Ref.~\cite{\BSQC}.
We refer to these two papers hereafter as BS2 and BS1, respectively.
This method involves two steps.
First, using time-ordered perturbation theory (TOPT), one derives
a form of the three-particle quantization condition containing an asymmetric,
 Lorentz-noninvariant three-particle K matrix. Here asymmetric refers to the fact that
 individual terms in this flavor matrix include only a subset of the contributing diagrams.
This ``asymmetric'' quantization condition is then converted to a form involving
a symmetric kernel (in which all diagrams are included) using symmetrization identities. 
The resulting kernel turns out also to be Lorentz invariant.
It is also possible, using the methods of BS2, to convert the asymmetric kernel into
Lorentz-invariant form.
In practice, the two forms of the formalism that are likely to be most useful are
that involving the symmetric kernel---for which the number of parameters is minimized---%
and the asymmetric but Lorentz-invariant form---in which we expect, following 
Refs.~\cite{\RtoK,\BSequiv}, that the kernel is related to the R matrix of Refs.~\cite{\Maiisobar,\isobar}.

Quantization conditions involving multiple flavors are most naturally written in terms of matrices
with flavor indices. For the nondegenerate theory, these matrices are three-dimensional,
with indices indicating the flavor of the spectator particle.
Here the spectator is the particle that does not participate when there is a two-particle interaction
(a definition made more precise below).
For the $2+1$ systems considered here, the analogous choice is to have two-dimensional flavor
matrices, corresponding to the two choices of spectator flavor.
This matches the matrix structure introduced in the study of the $DDK$ system in Ref.~\cite{\DDK}.

One might have thought one could obtain the formalism for $2+1$ systems by taking an
appropriate limit of the nondegenerate results. We have not, however, found a simple way to
take such a limit in the general case 
while accounting for the additional symmetry factors that are needed.
What we have achieved is an alternative derivation for a special case: the
$\pi^+\pi^+ D_s^\pm$ system in isosymmetric QCD. By isospin symmetry, this system
is equivalent to $(\pi^+\pi^0)_{I=2} D_s^\pm$, and this latter system can be treated using
the nondegenerate quantization condition of BS2, projected into the $I=2$ state of the two pions.
This serves as a nontrivial check on our main derivation.

To use our results in practice, one needs to have a parametrization of the three-particle 
K matrix. For nonresonant systems such as $\pi^+\pi^+ K^+$, a natural choice is the
threshold expansion~\cite{\BHSnum,\dwave}, which is
the analog of the effective range expansion for the two-particle K matrix.
Here we extend this expansion to both the nondegenerate and $2+1$ theories.

This article is organized as follows.
In Sec.~\ref{sec:2+1asymm} we derive the asymmetric form of the quantization
condition for $2+1$ systems, given in Eq.~(\ref{eq:QC3_2p1_asym}),
together with the relation between the intermediate
three-particle K matrix and the infinite-volume three-particle scattering amplitude.
Section~\ref{sec:2+1symm} contains the derivation of the symmetric form of the quantization condition
for $2+1$ systems, given in Eq.~(\ref{eq:QC3pipiK}),
as well as the relation of the symmetric three-particle K matrix to the physical
scattering amplitude. 
This is followed, in Sec.~\ref{sec:threshold},
by the description of the threshold expansions for the nondegenerate
and $2+1$ theories.
We end with conclusions and outlook in Sec.~\ref{sec:conc}.

We include two appendices. Appendix~\ref{sec:recap} contains a summary of the results
for three nondegenerate particles from BS2 and also defines notation.
Appendix~\ref{sec:2d+1} provides the above-described check on our results
by considering the limit of the fully nondegenerate formalism when two
of the particles are degenerate (but distinguishable)---a limit we refer to
as the ``$2d+1$" system. 

\section{Derivation of quantization condition with asymmetric $\Kdf$}
\label{sec:2+1asymm}

In this section we derive the asymmetric form of
the quantization condition for a system with two identical particles
(labeled $1$ and $1'$, with physical mass $m_1$)
and a third particle (labeled 2, with physical mass $m_2$) that is different.
This setup is similar to that obtained by setting $m_1=m_3$ in the
nondegenerate formalism summarized in Appendix~\ref{sec:recap},
but differs because here we assume that particles 1 and 3 are identical.
This leads to additional symmetry factors that percolate through the entire formalism.
To deal with these, we have found it simplest to
rederive the quantization condition from the beginning, 
although many of the steps can be carried over from the nondegenerate case with little change,
allowing the discussion to be significantly shortened.

We follow the approach laid out in BS2. We consider the correlation function
\begin{equation}
C^{2+1}_{3,L}(E,\bm P) = \int_L d^4x\, e^{i(E x^0 - \bm P \cdot \bm x)}
\langle 0 | T \sigma_{1 1' 2}(x) \sigma_{1 1' 2}^\dagger(0) | 0 \rangle\,,
\label{eq:C2p1def}
\end{equation}
where
$P^\mu =(E,\bm P)$ is the (fixed) total 4-momentum of the system and
$\sigma_{1 1' 2} \sim \phi_1^2 \phi_2$,
with $\phi_j$ the field that destroys particles of flavor $j$.
The subscript $L$ indicates
that space is confined to a cube of side length $L$, whereas the time integral is unconstrained.
We assume periodic boundary conditions in space, such that all spatial momenta (including $\bm P$)
are drawn from the finite-volume set: $\bm k = (2\pi/L) \bm n$, where $\bm n \in \mathbb Z^3$.

The range of validity of the formalism we develop is that for which
only the $2+1$ intermediate state  (with two particles of flavor 1 and one of flavor 2) can go on shell.
This constrains the allowed values of  the
center-of-momentum frame (CMF) energy, $E^*=\sqrt{E^2-\bm P^2}$, 
in a way that depends on the symmetries of the particles under consideration.
For example, if $m_1 < m_2$ and
there are separate $\phi_j\to -\phi_j$ symmetries associated with both fields,
then the range of validity is $m_2 < E^* < 4 m_1+m_2$.

The first step in the BS2 approach is to analyze the correlator $C^{2+1}_{3,L}(E,\bm P)$
using TOPT applied to the EFT describing the particles.
As a prelude to this step, we recall the corresponding results for 
both three nondegenerate particles and three identical particles.
We then recast the latter result in a form that makes the symmetry factors more
intuitive, allowing a straightforward generalization to the $2+1$ theory.
We find that the results can be compactly expressed using a matrix
notation involving two-dimensional matrices.
The final step is on-shell projection, which follows closely the approach of BS1
and BS2.

In the following, all undefined notation and terminology is explained in Appendix~\ref{sec:recap}.

\subsection{Recap of result for nondegenerate particles}
\label{subsec:nondegen}

The expression for the correlator for three distinguishable particles
(e.g.~with $\sigma \to \phi_1\phi_2\phi_3$), derived in BS2, is
\begin{align}
C_{3,L}^{\rm nd}(E,\bm P)
&= C_{3,\infty}^{{\rm nd},(0)}(E,\bm P) +
A' i D\frac1{1 -
 i(\overline\cB_{2,L}^{(1)} + \overline\cB_{2,L}^{(2)} +  \overline\cB_{2,L}^{(3)} +  \cB_3 ) i D} A\,.
\label{eq:CLnd}
\end{align}
Here $C^{\nd,(0)}_{3,\infty}$ (with ``nd" denoting ``nondegenerate") 
is the contribution that does not involve three-particle intermediate states, and which, 
for the kinematic range of interest, can be evaluated in infinite volume.\footnote{%
 More precisely, the difference between the finite-volume and infinite-volume results is exponentially suppressed in $L$.
We take such differences to be negligible throughout this paper,
and account explicitly only for power-law volume dependence.
}
The remainder of the expression contains kernels $A'$, $\overline{\cB}_{2,L}^{(i)}$, $\cB_3$,
and $A$, connected by cut factors $D$, all of which we now describe.

$\cB_3=\cB_3(E;\{\bm p\};\{\bm{k}\})$ is the sum over all connected,
amputated $3\to3$ TOPT diagrams that are three-particle irreducible in the s-channel (3PIs),
evaluated in infinite volume.
The initial-state momenta are specified by $\{\bm k\}\equiv \{\bm k_1,\bm k_2, \bm k_3\}$,
with $\bm k_i$ the momentum of the particle of flavor $i$.
The impact of working in finite volume is that the momenta take on discrete values
drawn from the finite-volume set, and thus serve as (implicitly summed) matrix indices.
They are constrained by $\sum_i \bm k_i = \bm P$,
so that labeling with all three momenta is redundant, 
but we do so as it shows the symmetry of the expressions.
The final-state momenta are similarly denoted $\{\bm p\}$.

The two-particle barred kernels, $\overline\cB_{2,L}^{(i)}$, are also matrices in $\{\bm k\}$ space,
and contain the infinite-volume 2PIs amplitudes $\cB_2^{(i)}$, 
which are labeled by the flavor $i$ of the 
spectator particle that does not take part in the interaction, and depend on the energy and
momenta of the two particles that scatter,
\begin{equation}
\overline\cB_{2,L}^{(i)}(E;\{\bm p \}; \{\bm k\})
 = 2\omega^{(i)}_{k_i} L^3 \delta_{\bm p_i \bm k_i} 
\cB_2^{(i)}(E-\omega^{(i)}_{k_i};\bm p_j,\bm p_\ell;\bm k_j,\bm k_\ell)\,.
\end{equation}
Here the flavors $i$, $j$, and $\ell$ are chosen to lie in cyclic order,
and $\omega^{(i)}_{k_i}$ is defined in Eq.~(\ref{eq:omegai}).

The quantities $A(E;\{\bm p\})$ and $A'(E;\{\bm k\})$ are, respectively,
the initial and final endcaps,
and result from the sum of all 3PIs TOPT diagrams connecting
the initial (final) operator to an amputated three-particle state.
These are also evaluated in infinite volume, and are, respectively,
row and column vectors in the matrix space.

The final quantity to describe is the matrix $D$, which connects the kernels.
It is associated with three-particle cuts (denoted ``relevant cuts'' in BS2) 
between adjacent kernels, and is given by
\begin{equation}
D_{\{\bm p\};\{\bm k\}}
= 
\delta_{\bm p_1 \bm k_1} \delta_{\bm p_2 \bm k_2} \delta_{\bm p_3 \bm k_3}
\frac1{L^6} \frac1{8 \omega^{(1)}_{k_1} \omega^{(2)}_{k_2}\omega^{(3)}_{k_3}} 
\frac1{E-\omega^{(1)}_{k_1}-\omega^{(2)}_{k_2}-\omega^{(3)}_{k_3}}
\,.
\label{eq:Dnd}
\end{equation}
An important distinction in the following is between F cuts and G cuts.
When one expands out the geometric series in Eq.~(\ref{eq:CLnd}),
F cuts arise when the factors of $D$ lie  between two-particle kernels
having the same spectator flavor
(e.g.~$\dots \cB^{(1)}_{2,L} D \cB^{(1)}_{2,L}\dots$),
while G cuts occur if the flavor changes
(e.g.~$\dots \cB^{(1)}_{2,L} D \cB^{(2)}_{2,L}\dots$).
If either (or both) of the adjacent kernels is a $\cB_3$, then the two types of cuts are
equivalent.

The final result we will need gives the relation between the correlator and the finite-volume
scattering amplitude $\cM_{3,L,\nd}^\off(E;\{\bm p\};\{\bm k\})$.
The latter is defined as the sum over all connected, amputated $3\to3$ TOPT diagrams, evaluated
in finite volume. The superscript indicates that it is, in general, off the energy shell,
i.e.~$\sum_i \omega^{(i)}_{k_i} \ne E\ne \sum_i \omega^{(i)}_{p_i}$.
It is conveniently packaged into the combination
\begin{align}
\cM_{23,L,\nd}^\off &= \sum_i \overline{\cM}_{2,L,\off}^{(i)} + \cM_{3,L,\nd}^\off\,,
\end{align}
where the finite-volume off-shell two-particle amplitudes are given by 
\begin{align}
\overline{\cM}_{2,L,\off}^{(i)} &= \overline{\cB}_{2,L}^{(i)} \frac1{1 - iD i\overline{\cB}_{2,L}^{(i)}}\,,
\end{align}
and where we have dropped energy and momentum arguments for brevity.
From BS2, we have the result
\begin{equation}
\cM_{23,L,\nd}^\off = 
(\overline\cB_{2,L}^{(1)} + \overline\cB_{2,L}^{(2)} +  \overline\cB_{2,L}^{(3)} +  \cB_3 ) 
\frac1{1 - i D
 i(\overline\cB_{2,L}^{(1)} + \overline\cB_{2,L}^{(2)} +  \overline\cB_{2,L}^{(3)} +  \cB_3 ) } \,,
 \label{eq:M23Lndoff}
\end{equation}
so that the correlator can be rewritten as
\begin{align}
C_{3,L}^{\rm nd}-
 C_{3,\infty}^{{\rm nd},(0)}
 =
A' i D A
+
A' iD i \cM_{23,L,\nd}^\off iD A\,.
\label{eq:CLtoM23Lnd}
\end{align}
An important implication of this result is that the quantization condition---determined by the
poles in $C_{3,L}^\nd$---can equally well be obtained from the poles in $\cM_{23,L,\nd}^\off$.

\subsection{New form for correlator for identical particles}
\label{subsec:idcorrelator}

As a useful step on the way to an expression for the correlator in the $2+1$ theory,
we recast the results for identical particles, obtained in BS1, in a form
that makes symmetry factors more explicit. 

We begin with a comment on notation.
We use the same definitions for $\overline \cB_{2,L}^{(i)}$ and $D$ as for nondegenerate particles,
except that here we use flavor labels $1$, $1'$, and $1''$, instead of $1$, $2$, and $3$,
in order to emphasize their identical nature.
In particular, the masses are equal: $m_1=m_{1'}=m_{1''}$.
The three-particle 3PIs kernel and the endcaps are also defined as for nondegenerate particles,
i.e.~as sums over the same classes of TOPT diagrams, 
and thus we use the same symbols.
This implies a different choice of normalization from that used in BS1:
\begin{equation}
\cB_3 = 9 \cB_3^\text{BS1}\,,\quad
A' = 3 A'_\text{BS1}\,,\quad
A = 3 A_\text{BS1}\,.
\label{eq:equivalences}
\end{equation}
Another change in notation compared to BS1 is that here we use all three
momenta as matrix indices (as for nondegenerate particles), whereas in BS1
only the two independent momenta are used.

As will be shown below, the correlator for three identical particles 
(obtained with $\sigma\to \phi_1^3$, for example) 
can be written
\begin{align}
C_{3,L}^\ID - C_{3,L}^{\ID,(0)} &=
A' \frac{iD}{3!} \frac1{1 - i \left( \cS_\ID \overline \cB_{2,L}^{(1)} \cS_\ID + \cB_3 \right) 
\frac{iD}{3!} } A\,,
\label{eq:C3LIDnew}
\end{align}
where we have introduced a symmetrization operator
\begin{equation}
\cS_\ID \equiv 1 + P^{(11')} + P^{(1 1'')}\,.
\label{eq:SID}
\end{equation}
Here $P^{(11')}$ interchanges the indices $1$ and $1'$
\begin{equation}
P^{(11')}_{\{\bm p\};\{\bm k\}} \equiv \delta_{\bm p_1 \bm k_{1'}}
\delta_{\bm p_{1'} \bm k_{1}} \delta_{\bm p_{1''} \bm k_{1''}}\,,
\end{equation}
while $P^{(11'')}$ interchanges $1$ and $1''$.
The key features of the new form, Eq.~(\ref{eq:C3LIDnew}),
are that the symmetry factor $3!$ is explicitly associated with each relevant cut,
and that the two-particle kernel is symmetrized.\footnote{%
Note that, to symmetrize, one needs only three contributions in $\cS_\ID$, rather than $3!=6$, 
because the two-particle kernel is already symmetric under the interchange of the two particles that
are scattering.}

To demonstrate that Eq.~(\ref{eq:C3LIDnew}) is correct, one can either derive it directly from a 
diagrammatic analysis, or show that it leads to the same result as that obtained in BS1.
We follow the latter path here. Key results that we need are that $\cB_3$, $A_\ID$,
and $A'_\ID$ are invariant under particle exchange, e.g.
\begin{equation}
P^{(11')} \cB_3 = P^{(11'')} \cB_3 = \cB_3
= \cB_3 P^{(11')}= \cB_3 P^{(11'')}\,.
\end{equation}
This implies that
\begin{equation}
\cS_\ID \frac{\cB_3}9 \cS_\ID = \cB_3\,,\quad
A' = \frac{A'}3 \cS_\ID\,,\quad
A = \cS_\ID \frac{A}3 \,,
\end{equation}
which allows us to rewrite Eq.~(\ref{eq:C3LIDnew}) as
\begin{align}
C_{3,L}^\ID - C_{3,L}^{\ID,(0)} &=
\frac{A'}3 \cS_\ID \frac{iD}{3!} 
\frac1{1 -  \cS_\ID i(\overline \cB_{2,L}^{(1)} + \frac{\cB_3}9) \cS_\ID
\frac{iD}{3!} } \cS_\ID \frac{A}{3}
\label{eq:C3LIDa}
\\
&=
\frac{A'}3 \cS_\ID \frac{iD}{3!} \cS_\ID
\frac1{1 -  i(\overline \cB_{2,L}^{(1)} + \frac{\cB_3}9) \cS_\ID
\frac{iD}{3!}  \cS_\ID }\frac{A}{3}\,.
\label{eq:C3LIDb}
\end{align}
In words, we have transferred the symmetrization from the kernels
onto the cut factor $D$.
The final step is to note that
$\cS_\ID D \cS_\ID$ contains two types of contribution:
three diagonal terms, with the same factor from $\cS_\ID$ on both sides,
e.g.~$P^{(11')} D P^{(11')}$, 
and six off-diagonal terms, e.g.~$D P^{(11')}$.
Since $D$ is symmetric when the particle masses are equal
[see Eq.~(\ref{eq:Dnd})],
the diagonal terms are the same and equal $D$.
The off-diagonal terms are not all equal, but, when adjacent to a kernel
that is symmetric under $1'\leftrightarrow 1''$ interchange (which holds for all the
kernels in $C_{3,L}^\ID$), they can all be brought into the form $D P^{(11')}$.
Thus we find 
\begin{equation}
\cS_\ID \frac{D}{3!} \cS_\ID \to \frac{D}{2} + D P^{(11')}
\equiv D_F^\ID + D_G^\ID\,,
\label{eq:DID}
\end{equation}
where in the second step we have used the definitions from BS1,
which include a factor of $1/2$ in $D_F^\ID$.
Substituting this result into Eq.~(\ref{eq:C3LIDb}),
and using the equivalences in Eq.~(\ref{eq:equivalences}),
we obtain Eq.~(24) of BS1,
thus confirming the validity of Eq.~(\ref{eq:C3LIDnew}).

When lying between two $\overline B_{2,L}^{(1)}$ kernels,
the two terms in Eq.~(\ref{eq:DID}) correspond to F and G cuts.
In the former, both two-particle scatterings involve the same pair,
while in the latter, due to the presence of $P^{(11')}$, the scattering pair changes.

The finite-volume three-particle scattering amplitude for identical particles,
$\cM_{3,L,\ID}^\off$, can also
be rewritten using the same building blocks as Eq.~(\ref{eq:C3LIDnew}).
We find that
\begin{align}
\cM_{23,L,\ID}^\off &\equiv \cS_\ID \overline \cM_{2,L,\off}^{(1)} \cS_\ID + \cM_{3,L,\ID}^\off
\\
&=\left( \cS_\ID \overline \cB_{2,L}^{(1)} \cS_\ID + \cB_3^\ID \right)
\frac1{1 - \frac{i D}{3!} i \left( \cS_\ID \overline \cB_{2,L}^{(1)} \cS_\ID + \cB_3^\ID \right)}\,,
\label{eq:M23LID}
\end{align}
where 
\begin{align}
\overline{\cM}_{2,L,\off}^{(1)} &= \overline{\cB}_{2,L}^{(1)} 
\frac1{1 - iD_F^\ID i\overline{\cB}_{2,L}^{(1)}}\,.
\end{align}
The symmetry-related factors in this result have an intuitively reasonable form.
The result (\ref{eq:M23LID}) shows that the full $\cM_{3,L}$ can be written as a geometric series
as long as it is combined with the symmetrized two-particle amplitude.
This was not realized in BS1, where a simple form was only
obtained for an unsymmetrized version of the amplitude.

To show that Eq.~(\ref{eq:M23LID}) is correct, we rewrite it as
\begin{align} 
\cM_{23,L,\ID}^\off 
&=\cS_\ID \bigg( \overline \cB_{2,L}^{(1)} + \frac{\cB_3^\ID}9 \bigg)
\frac1{1 - \cS_\ID \frac{iD}{3!} \cS_\ID i\Big( \overline \cB_{2,L}^{(1)} + \frac{\cB_3^\ID}9\Big)} \cS_\ID\,.
\label{eq:M23LIDb}
\end{align}
which, using Eq.~(\ref{eq:DID}), is equivalent to the
result given in Eqs.~(83) and (84) of BS1.

The final new result we present is the relation between the correlation function and
the scattering amplitude,
\begin{align}
C_{3,L}^\ID-
 C_{3,\infty}^{\ID,(0)}
 =
A'  \frac{i D}{3!} A
+
A' \frac{iD}{3!} i \cM_{23,L,\ID}^\off \frac{iD}{3!} A\,.
\label{eq:CLtoM23LID}
\end{align}
which has a similar form to the corresponding result for nondegenerate particles, 
Eq.~(\ref{eq:CLtoM23Lnd}).

\subsection{Results for the 2+1 theory}
\label{subsec:2p1correlator}

We now return to the $2+1$ theory.
We take the matrix indices to be
$\{\bm k\}=\{\bm k_1,\bm k_{1'},\bm k_2\}$,
with $m_{1'}=m_1$.
By generalizing the approach of the previous two subsections, we find
that the correlator, defined in Eq.~(\ref{eq:C2p1def}), is given by
\begin{align}
C_{3,L}^{2+1} - C_{3,\infty}^{2+1,(0)}
&=
A' \frac{iD}2 \frac1{1 - i \left(\cS_{11'}\overline \cB_{2,L}^{(1)} \cS_{11'}
+ \overline \cB_{2,L}^{(2)} + \cB_3 \right) \frac{iD}2 } A\,,
\label{eq:C3L2p1}
\end{align}
where the symmetrization operator is
\begin{equation}
\cS_{11'} = 1 + P^{(11')}\,,
\end{equation}
and the symmetry factor associated with the cut is now $1/2!$.
The validity of this expression can be shown by a straightforward diagrammatic
analysis, and we do not present the details.
We stress that the kernels $A'$, $A$, $\cB_3$,
and $\overline\cB_{2,L}^{(i)}$ are defined by the same sum of TOPT diagrams
as in the earlier discussion of the nondegenerate and identical cases,
except here two of the particles are identical. Thus we continue to use the same
symbols for these kernels.

A similar analysis leads to the following result for the finite-volume three-particle
scattering amplitude, $\cM_{3,L,2+1}^\off$,
\begin{align}
\cM_{23,L,2+1}^\off &\equiv 
\cS_{11'} \overline \cM_{2,L,\off}^{(1)} \cS_{11'} + \overline \cM_{2,L,\off}^{(2)}  +
\cM_{3,L,2+1}^\off 
\label{eq:M23L2p1a}
\\
&=\left(\cS_{11'}\overline \cB_{2,L}^{(1)} \cS_{11'}
+ \overline \cB_{2,L}^{(2)} + \cB_3 \right)
\frac1{1 - \frac{i D}{2} i 
\left(\cS_{11'}\overline \cB_{2,L}^{(1)} \cS_{11'}
+ \overline \cB_{2,L}^{(2)} + \cB_3 \right)} \,,
\label{eq:M23L2p1}
\end{align}
where as usual we have added in parts containing only two-particle interactions, 
given here by
\begin{align}
\overline \cM_{2,L,\off}^{(1)}  &=
 \overline \cB_{2,L}^{(1)} \frac1{1 - iD i \overline \cB_{2,L}^{(1)}}\,,
 \label{eq:M2L12p1}
 \\
\overline \cM_{2,L,\off}^{(2)}  &=
 \overline \cB_{2,L}^{(2)} \frac1{1 - \frac{iD}2 i \overline \cB_{2,L}^{(2)}}\,.
 \label{eq:M2L22p1}
\end{align}
The factor of $1/2$ in (\ref{eq:M2L22p1}) is the symmetry factor
in loops containing flavors $1$ and $1'$.
As above, we can also express the correlator in terms of this amplitude
\begin{align}
C_{3,L}^{2+1}-
 C_{3,\infty}^{2+1,(0)}
 =
A'  \frac{i D}{2} A
+
A' \frac{iD}{2} i \cM_{23,L,2+1}^\off \frac{iD}{2} A\,.
\label{eq:CLtoM23L2p1}
\end{align}

\subsection{Matrix form of results for $2+1$ theory}
\label{subsec:matrix}

The next stage in the derivation is to project the kernels adjacent to the
three-particle cuts on shell, using the relations given in BS2
(which themselves are generalizations of those in Refs.~\cite{\HSQCa} and BS1).
These on-shell projections single out one of the momenta on each side of the cut as the spectator,
and their forms depend on whether the same particle is the spectator on both sides of the cut
(if yes, then an F cut is used; if no, a G cut),
as well as on the masses of the spectator(s).

In order to keep track of which projection appears for each cut, 
it is useful to introduce an additional layer of flavor matrix indices,
following the approach of BS2 for the nondegenerate theory.
We distinguish matrix versions of quantities using carets (``hats'').
There is considerable freedom in how the matrix forms are written,
and in the choice of dimensions of the matrices.
In particular, one can use a three-dimensional form as in BS2,
or a two-dimensional form corresponding to the two types of particles.
We have worked out both choices, but present only the latter as it will be simpler to
implement in practice.

The first step is to rewrite the results of the previous subsection in a way that
makes clear whether cuts lying between factors of $\overline\cB_{2,L}^{(1)}$ are of
F or G type. Following Eq.~(\ref{eq:DID}), we define
\begin{equation}
\begin{split}
D_F^{2+1} &= D = P^{(1 1')}  D P^{(1 1')} \,,
\\
D_G^{2+1}  &= P^{(1 1')} D = D P^{(1 1')} \,.
\end{split}
\label{eq:DFG2p1}
\end{equation}
We stress that there is no factor of $1/2$ in $D_F^{2+1}$, unlike in $D_F^\ID$.
We next move the action of the symmetrization operators from the kernels
onto the factors of $D$, following the approach used for identical particles.
To do so, we need the symmetry relations
\begin{equation}
\begin{split}
P^{(11')} \cB_3 &= \cB_3 = \cB_3 P^{(11')}\,,
\\
P^{(11')} \overline{\cB}^{(2)}_{2,L} &= \overline{\cB}^{(2)}_{2,L}  
= \overline{\cB}^{(2)}_{2,L} P^{(11')}\,,
\\
A' P^{(11')} &= A'\,,\quad P^{(11')} A = A\,,
\end{split}
\label{eq:2p1symms}
\end{equation}
which follow from the identical nature of particles $1$ and $1'$.
Using these, we have
\begin{equation}
A' = \frac{A'}2\cS_{11'}\,,\quad
A = \cS_{11'} \frac{A}2\,,
\end{equation}
and
\begin{equation}
\cS_{11'}\overline \cB_{2,L}^{(1)} \cS_{11'} + \overline \cB_{2,L}^{(2)} + \cB_3 
=
\cS_{11'} \left(\overline \cB_{2,L}^{(1)} 
+ \tfrac14\overline \cB_{2,L}^{(2)} + \tfrac14 \cB_3 \right) \cS_{11'}\,.
\end{equation}
This allows us to rewrite Eq.~(\ref{eq:C3L2p1}) as
\begin{align}
C_{3,L}^{2+1} - C_{3,\infty}^{2+1,(0)}
&=
\frac{A'}2  \cS_{11'} \frac{iD}2 \cS_{11'} \frac1{1 -
i( \overline\cB_{2,L}^{(1)} + \tfrac14 \overline\cB_{2,L}^{(2)} + \tfrac14 \cB_3 )
 \cS_{11'} \frac{iD}2 \cS_{11'} } 
\frac{A}2
\\
&=
\frac{A'}2 i (D_F^{2+1}+D_G^{2+1})\frac1{1 -
i( \overline\cB_{2,L}^{(1)} + \tfrac14 \overline\cB_{2,L}^{(2)} + \tfrac14 \cB_3 )
 i (D_F^{2+1}+D_G^{2+1})} 
\frac{A}2 \,,
\label{eq:CL2p1}
\end{align}
where to obtain the second form we have used
\begin{equation}
\cS_{11'} \frac{D}2 \cS_{11'} =D_F^{2+1}+D_G^{2+1}\,,
\end{equation}
which follows from the definitions in Eq.~(\ref{eq:DFG2p1}).

We are now ready to convert to a matrix form.
As noted above, there are many ways to do this, and we make a choice that 
allows the most direct comparison with the results
obtained in Appendix~\ref{sec:2d+1} from a projection of the nondegenerate formalism.
We introduce the two-dimensional vectors\footnote{%
The factors of $1/\sqrt2$ appearing in $\bra\alpha$ and $\ket\alpha$ are chosen so as
to simplify the comparison with the results of Appendix~\ref{sec:2d+1}.}
 \begin{equation}
\bra\alpha =(1, 1/\sqrt2)\ \ {\rm and} \ \
\ket\alpha= \begin{pmatrix} 1 \\ 1/\sqrt2 \end{pmatrix}\,,
\label{eq:alphadef}
 \end{equation}
 in terms of which we write a matrix form of the cut factor 
 \begin{equation}
\wh  D = \ket{\alpha} (D_F^{2+1} + D_G^{2+1}) \bra{\alpha}\,.
\label{eq:Dmat2p1}
\end{equation}
To avoid overloading the notation
we use the same names for matrix quantities as in Appendix~\ref{sec:recap}, 
although here they are two-dimensional. 
The matrix form of Eq.~\eqref{eq:CL2p1} is then
\begin{align}
C_{3,L}^{2+1} - C_{3,\infty}^{2+1,(0)} &=
\wh A' i\wh D \frac1{1- i  \wh\cB i \wh D} \wh A \,,
\label{eq:C3L2p1mat}
\end{align}
where $\hat A'$ is a row vector, $\wh \cB$ is a matrix, and $\wh A$ is a column vector. 
To demonstrate the equivalence of Eqs.~\eqref{eq:CL2p1} and (\ref{eq:C3L2p1mat})
we insert the matrix form of $\wh D$ into the latter, so that the right-hand side becomes
\begin{align}
\wh A' \ket{\alpha} (D_F^{2+1} + D_G^{2+1})\bra{\alpha} 
\frac1{1 +  \wh\cB   \ket{\alpha} (D_F^{2+1} + D_G^{2+1})\bra{\alpha} } \wh A 
=
\wh A' \ket{\alpha} (D_F^{2+1} + D_G^{2+1}) 
\frac1{1 +  \bra\alpha \wh\cB \ket\alpha (D_F^{2+1} + D_G^{2+1})}\bra{\alpha}  \wh A \,.
\end{align}
This agrees with Eq.~\eqref{eq:CL2p1} as long as
\begin{equation}
\bra\alpha \wh \cB \ket \alpha = 
\overline\cB_{2,L}^{(1)} + \tfrac14 \overline\cB_{2,L}^{(2)} + \tfrac14 \cB_3 \,,\quad
\wh A'\ket{\alpha} = A'\,, \ \ {\rm and} \ \
\bra{\alpha}\wh A = A \,.
\end{equation}
We can satisfy the constraint on $\wh \cB$ with the following choices:
\begin{align}
	\wh\cB &= \wh{\overline{\cB}}_{2,L} + \wh\cB_3 \,,
\label{eq:Bmat}
	\\
	\wh{\overline{\cB}}_{2,L} &= 
{\rm diag}\left(\overline\cB_{2,L}^{(1)},\tfrac12 \overline\cB_{2,L}^{(2)} \right) \,,
\label{eq:B2Lmat}
\\
\bra\alpha \wh \cB_3 \ket\alpha &=  \tfrac14 \cB_3 \,.
\label{eq:B3mat}
\end{align}
The constraints on $\wh\cB_3$, $\wh A'$, and $\wh A$ leave considerable freedom in the
matrix forms, but we will not need to make specific choices.
The freedom in choosing these quantities is related to the ambiguity in defining asymmetric kernels,
a point discussed in BS1 and BS2.

The form of $\wh{\overline{\cB}}_{2,L}$ in Eq.~(\ref{eq:B2Lmat}) is chosen to allow,
to the extent possible, a clear separation between F and G cuts.
We know from the symmetry relations (\ref{eq:2p1symms}), together with the
definitions of $D_F^{2+1}$ and $D_G^{2+1}$ in Eq.~(\ref{eq:DFG2p1}),
that these two cut factors are equivalent when adjacent to endcaps or factors of $\cB_3$
or $\overline\cB_{2,L}^{(2)}$. This allows us to rewrite $\wh D$ in Eq.~(\ref{eq:C3L2p1mat}) as
\begin{align}
\wh D &\to \wh D_F + \wh D_G\,,
\label{eq:Dhat2p1_FpG}
\\
\wh D_F &= {\rm diag}\left( D_F^{2+1}, D_F^{2+1}\right)\,,
\label{eq:DFhat}
\\
\wh D_G &=
\begin{pmatrix}
D_G^{2+1} & \sqrt 2 D_G^{2+1} \\ \sqrt 2 D_G^{2+1} &  0
\end{pmatrix}\,.
\label{eq:DGhat}
\end{align}
This form is chosen since the off-diagonal entries will become G cuts, 
as they necessarily involve a change of spectator if lying between two-particle kernels, 
while the bottom-right entry will become an F cut.

By similar manipulations, we can write $\cM_{23,L,2+1}^\off$, 
given by Eq.~(\ref{eq:M23L2p1}), in matrix form.
The result is
\begin{align}
\cM_{23,L,2+1}^\off &= \cS_{11'} \bra\alpha \wh \cM_{23,L,2+1}^\uu \ket\alpha \cS_{11'}\,,
\label{eq:M23L2p1b}
\\
\wh \cM_{23,L,2+1}^\uu &= \wh \cB \frac1{1 -i \wh D i\wh \cB}\,,
\label{eq:M23L2p1uu}
\end{align}
where the second form of $\wh D$, Eq.~(\ref{eq:Dhat2p1_FpG}), can also be used.
Here the superscript ``$\uu$'', adapted from Ref.~\cite{\HSQCa},
indicates that the quantity is unsymmetrized, 
in the sense that not all diagrams are included in each entry of the matrix.

\subsection{On-shell projection}
\label{subsec:onshell}

We are now ready to project the kernels and endcaps on shell, 
so as to obtain a form for the correlator and finite-volume scattering amplitude
that is written in terms of on-shell quantities, which can, in turn, be related to observables.
We use the method explained in BS1 and generalized in BS2.
As discussed in Appendix~\ref{sec:recap}, this converts the off-shell indices $\{\bm k\}$ into the
on-shell set $\{k_i\ell m i \}$. As indicated by the presence of the index $i$,
the choice of coordinate system depends on which of
the three momenta is treated as the spectator.
The choice made depends on the position in the flavor matrix.

For each choice of spectator flavor,
there is an additional choice as to which of the nonspectator pair is used to decompose
into spherical harmonics.
We choose a different convention from the cyclic one used for nondegenerate particles
and described in Appendix~\ref{sec:recap}.
First we note that, in our two-dimensional flavor setup, the spectator can have flavor
$1$ or $2$ (but not $1'$).
Our convention here is that the particle whose direction (in the pair CMF)
is used to define the spherical harmonics {\em always has flavor $1'$}.
In other words, the cyclic order is replaced by $\{1,1',2\}$ and $\{2,1',1\}$.
This applies to the kernels, endcaps, and the kinematic function $\wt F^{(i)}$,
but not for $\wt G^{(ij)}$, where the definition (\ref{eq:Gt}) completely specifies the particle
directions used in the spherical harmonics.
We note that, if the spectator has flavor $2$,
it does not matter whether we choose flavor $1$ or $1'$ to define the harmonics,
because the adjacent kernels are symmetric under $1\leftrightarrow 1'$ interchange.
In particular this latter property implies that only even angular momenta appear in the
corresponding expansion in spherical harmonics for kernels with flavor $2$ as the spectator.

On-shell projection is achieved by writing\footnote{%
As in the previous subsection, we continue here to lighten the notation
by using the same names for matrices---$\wh F_G$,
$\wh F$, $\wh G$, etc.---as in Appendix~\ref{sec:recap}, although the two sets of matrices
are different, here being two-dimensional whereas in the appendix they are three-dimensional.}
\begin{equation}
\wh D_F = \wh F + \wh{\delta F}\,,\quad
\wh D_G = \wh G + \wh{\delta G}\,,
\end{equation}
so that
\begin{align}
\wh D &= \wh{F}_G + \wh{\delta F_G}\,, 
\label{eq:Dhat2p1}
\\
\wh F_G &= \wh F + \wh G\,, 
\label{eq:FG2p1}
\\
\wh{\delta F_G}&=\wh{\delta F}+\wh{\delta G}\,,
\label{eq:deltaFG2p1}
\end{align}
with
\begin{align}
\wh F &= {\rm diag}\left( \wt F^{(1)}, \wt F^{(2)} \right)\,,
\label{eq:Fhat2p1}
\\
\wh G &=
\begin{pmatrix}
\wt G^{(11)} & \sqrt2 P_L \wt G^{(12)}\\
\sqrt2 \wt G^{(21)}P_L & 0
\end{pmatrix}
\,,
\label{eq:Ghat2p1}
\end{align}
Here $\wt F^{(i)}$, $\wt G^{(ij)}$, and $P_L$ are defined in Eqs.~(\ref{eq:Ft}), (\ref{eq:Gt}),
and (\ref{eq:PLdef}), respectively, aside from the above-mentioned change in the order
of flavor indices for $\wt F^{(i)}$, and the introduction of $\wt G^{(11)} \equiv \wt G^{(11')}$.
The factors of $P_L$ are needed because the convention for defining spherical harmonics
in $\wt G^{(12)}$ and $\wt G^{(21)}$ differs from that defined above that is used for the kernels.

In the decomposition (\ref{eq:Dhat2p1}), any kernel adjacent to $\wh{F}_G$ is set on shell.
The remainder, $\wh{\delta F_G}$,
which is finite at the poles in $\wh D_F$ and $\wh D_G$,
can be absorbed into the infinite-volume kernels by rearranging the geometric series.
The correlator (\ref{eq:C3L2p1mat}) becomes
\begin{align}
C_{3,L}^{2+1}- C_{3,\infty}^{2+1,(0)}
&= \wh A^{\prime(u)} i\wh{F}_G
\frac1{1 + \wh \cK_{\df,23,L}^\uu \wh{F}_G} \wh A^{(u)}\,, 
\label{eq:C3L_2p1_on}
\end{align}
where
\begin{align}
	\wh \cK_{\df,23,L}^\uu
	&= \wh \cB \frac1{1 + \wh{\delta F_G} \wh \cB} \Bigg|_{\text{on shell}}\,,   
\label{eq:Kdf23L_2p1}
\\
\wh A^{\prime(u)} &= \wh A' \frac1{1 + \wh{\delta F_G} \wh \cB} \Bigg|_{\text{on shell}}\,,   
\\
\wh A^{(u)} &=  \frac1{1 + \wh \cB \wh{\delta F_G}} \wh A\Bigg|_{\text{on shell}}    
\label{eq:Auhat2p1}
\end{align}
are on-shell, infinite-volume quantities.
As discussed in Appendix~\ref{sec:recap}, the form of the on-shell projection is
different for each element of the matrices, because a different spectator flavor is used.
The ``$u$'' superscripts on kernels again indicate that they are asymmetric, i.e.~that
each element of the matrices is obtained by summing a different subset of diagrams.

As in BS2, we can separate the two- and three-particle parts of
the interaction kernel,
\begin{equation}
	\wh \cK_{\df,23,L}^\uu \equiv 
	\wh{\overline{\cK}}_{2,L} + \wh\cK_{\df,3}^\uu \,,   
	\label{eq:Kdf23Lhat2p1}
\end{equation}
with
\begin{align}
	\wh{\overline{\cK}}_{2,L} &= 
	\wh{\overline{\cB}}_{2,L} \frac1{1+\wh{\delta F} 	\wh{\overline{\cB}}_{2,L}}\Bigg|_{\text{on shell}}
	\\
	&= \text{diag} \left(\overline{\cK}_{2,L}^{(1)} \,,  \tfrac12\overline{\cK}_{2,L}^{(2)} \right) \,,
	\label{eq:K2Lhat2p1}
\end{align}
where $\overline{\cK}_{2,L}^{(i)}$, defined in Eq.~(\ref{eq:K2Li}), contains 
on-shell two-particle K matrices. 
We note that this construction automatically includes the correct
intermediate symmetry factors of $1$ for $\overline{\cK}_{2,L}^{(1)}$ and $1/2$
for $\overline{\cK}_{2,L}^{(2)}$.
We stress that the quantity $\wh\cK_{\df,3}^\uu$ is an infinite-volume amplitude,
as can be shown using essentially the same arguments as given in BS1 and BS2.

We can also project the finite-volume scattering amplitude on shell using similar 
manipulations. For now,
we consider the asymmetric version, given in Eq.~(\ref{eq:M23L2p1uu}),
whose on-shell projection can be written
\begin{align}
\wh \cM_{23,L}^{\uu}\Big|_{\text{on shell} }
&= \wh \cK_{\df,23,L}^\uu \frac1{1+ \wh F_G \wh \cK_{\df,23,L}^\uu} \,.
\label{eq:M23Luuhat}
\end{align}

\subsection{Quantization condition and relation of $\wh\cK_{\df,3}^\uu$ to $\cM_3$}
\label{subsec:asymQC}

From the form of the correlator in Eq.~(\ref{eq:C3L_2p1_on}) 
[or from Eq.~(\ref{eq:M23Luuhat})] we can read off the
asymmetric version of the quantization condition,
\begin{equation}
\det\left(1 + \wh{F}_G [\wh{\overline{\cK}}_{2,L} + \wh\cK_{\df,3}^\uu] \right) = 0\,.
\label{eq:QC3_2p1_asym}
\end{equation}
This has the same form as that in the nondegenerate theory, 
Eq.~(\ref{eq:QC3ndasym}),
but differs as the matrices here are two-dimensional and contain symmetry factors.
The latter are the factor of $1/2$ in $\wh{\overline{\cK}}_{2,L}$, Eq.~(\ref{eq:K2Lhat2p1}),
and the $\sqrt 2$ in the off-diagonal terms in $\wh G$, Eq.~(\ref{eq:Ghat2p1}).

The three-particle amplitude that appears in the quantization condition,
$\wh \cK_{\df,3}^\uu$, while an infinite-volume quantity, is unphysical.
This is because it depends on the cutoff functions $H^{(i)}(\vec k)$ contained in $\wh F_G$
[see Eqs.~(\ref{eq:Ft}) and (\ref{eq:Gt})],
and also upon the principal value scheme used to define integrals over poles.
For the quantization condition to be useful, one must, in a second step, relate
$\wh \cK_{\df,3}^\uu$ to the physical scattering amplitude. The method for doing so
was developed in Ref.~\cite{\HSQCb}, and applied to the asymmetric quantization condition
in BS1 and BS2. We follow the same path here.

We begin by applying on-shell projection to the amplitude $\cM_{3,L,2+1}^\off$.
Using Eqs.~(\ref{eq:M23L2p1a}), (\ref{eq:M23L2p1b}),
and the symmetry of $\overline\cM_{2,L,\off}^{(2)}$, this can be written as
\begin{align}
\cM_{3,L,2+1}^\off &= \cS_{11'} \left(
\bra\alpha \wh \cM_{23,L,2+1}^\uu\ket\alpha 
-\overline\cM_{2,L,\off}^{(1)} -\tfrac14 \overline\cM_{2,L,\off}^{(2)}
\right) \cS_{11'} 
\label{eq:M3L2p1offb}
\\
&\equiv \cS_{11'} \bra\alpha \wh \cM_{3,L,2+1}^\uu \ket\alpha \cS_{11'} \,,
\end{align}
where
\begin{align}
\wh\cM_{3,L,2+1}^\uu &\equiv
\wh \cM_{23,L,2+1}^\uu - \wh{\overline{\cM}}_{2,L,\off} \,,
\\
\wh{\overline{\cM}}_{2,L,\off} &= {\rm diag}\left(
\overline\cM_{2,L,\off}^{(1)}, \tfrac12 \overline\cM_{2,L,\off}^{(2)}\right)\,,
\label{eq:M2Lhat2p1}
\end{align}
with the $\overline\cM_{2,L,\off}^{(i)}$ given in Eqs.~(\ref{eq:M2L12p1}) and (\ref{eq:M2L22p1}).
Note that the $1/4$ in Eq.~(\ref{eq:M3L2p1offb}) changes to a $1/2$ in Eq.~(\ref{eq:M2Lhat2p1})
because of the factors of $1/\sqrt2$ in the definitions of $\bra\alpha$ and $\ket\alpha$.
We first apply on-shell projection to $\wh\cM_{3,L,2+1}^\uu$, which, using
Eq.~(\ref{eq:M23Luuhat}), yields
\begin{equation}
\wh\cM_{3,L,2+1}^\uu\Big|_{\text{on shell}} =
 \wh \cK_{\df,23,L}^\uu \frac1{1+ \wh F_G \wh \cK_{\df,23,L}^\uu} - 
 \wh{\overline{\cM}}_{2,L}\,,
 \label{eq:M3L2p1uuhat}
\end{equation}
where $\wh{\overline{\cM}}_{2,L}$, the on-shell projection of $\wh{\overline{\cM}}_{2,L,\off}$,
is related to the corresponding quantity containing K matrices by
\begin{equation}
\wh{\overline{\cM}}_{2,L}^{-1} =\wh{\overline{\cK}}_{2,L}^{-1} + \wh F
\,.
\label{eq:M2Lhat2p1on}
\end{equation}
The on-shell projection of $\cM_{3,L,2+1}^\off$ is then given by
\begin{align}
\cM_{3,L,2+1} &= \bra{\alpha_\cS} \wh\cM_{3,L,2+1}^\uu\Big|_{\text{on shell}} \ket{\alpha_\cS}\,,
\label{eq:M3L2p1final}
\end{align}
where $\bra{\alpha_\cS}\sim \cS_{11'} \bra\alpha$ and $\ket{\alpha_\cS}\sim \ket\alpha \cS_{11'}$.
We use the symbol ``$\sim$'' because of an additional subtlety. The elements of
the on-shell projection of the matrix $\wh\cM_{3,L,2+1}^\uu$ are in the $\{k_i \ell m i\}$ basis.
To be added and symmetrized, they must first be combined with the appropriate spherical harmonics,
as in Eq.~(\ref{eq:Xdecomp}), so that they are a function of three on-shell momenta.
This step is implicitly included in the definitions of $\bra{\alpha_\cS}$ and $\ket{\alpha_\cS}$.\footnote{%
In fact, it is simplest to take the infinite-volume limit discussed in the next paragraph first,
and then combine with spherical harmonics. This is because momenta are then continuous,
so one does not have to deal with subtleties arising from the fact that the different
choices of on-shell projection lead to slightly different on-shell triplets of momenta.
}

We now perform some rather involved algebraic manipulations so as
to make the dependence of $\wh\cM_{3,L,2+1}^\uu$ on $\wh\cK_{\df,3}^\uu$ explicit.
These steps, which start from Eq.~(\ref{eq:M3L2p1uuhat}), 
and use Eqs.~(\ref{eq:FG2p1}), (\ref{eq:Kdf23Lhat2p1}), and (\ref{eq:M2Lhat2p1on}), 
are identical to those followed in 
BS1 and BS2 (see Appendix C of the former work), and lead to
\begin{align}
\wh \cM_{3,L,2+1}^{(u,u)}\Big|_{\text{on shell}} 
&=  \wh \cD_L^{(u,u)}+  \wh \cM_{\df,3,L}^{(u,u)}\,,
\label{eq:M3Lhatdecomp}
\\
\wh \cD_L^{(u,u)} &= - \wh{\overline{\cM}}_{2,L} \wh G \wh{\overline{\cM}}_{2,L}
\frac1{1 +\wh G  \wh{\overline{\cM}}_{2,L} }\,,
\label{eq:DLhat}
\\
 \wh \cM_{\df,3,L}^{(u,u)} &=
\left[1- \wh \cD_{23,L}^\uu  \wh F_G\right]
 \wh \cK_{\df,3}^\uu
\frac1{ 1 + [1 - \wh F_{G}  \wh \cD_{23,L}^\uu]  \wh F_G  \wh \cK_{\df,3}^\uu}
\left[1- \wh F_G  \wh \cD_{23,L}^\uu\right]\,,
\label{eq:Mdf3Lhat}
\\
\wh \cD_{23,L}^{(u,u)} &=  \wh{\overline{\cM}}_{2,L} + \wh \cD_L^{(u,u)}
=  \wh{\overline{\cM}}_{2,L}
\frac1{1+ \wh G  \wh{\overline{\cM}}_{2,L} }\,.
\label{eq:D23Lhat}
\end{align}
These results have exactly the same form as in the nondegenerate theory,
except here the matrices are two-dimensional.
Following the approach of Ref.~\cite{\HSQCb}, one can now take the $L\to\infty$ limit
in an appropriate manner, such that $\cM_{3,L,2+1}$ goes over to the infinite-volume
scattering amplitude for the $2+1$ theory, $\cM_{3,2+1}$.
As described in detail in BS2, this limit converts the results in Eqs.~(\ref{eq:DLhat})-(\ref{eq:D23Lhat})
into integral equations. Solving these integral equations,
and substituting the results into Eq.~(\ref{eq:M3Lhatdecomp}), and thence into
Eq.~(\ref{eq:M3L2p1final}), one obtains $\cM_{3,2+1}$ in terms of the elements
of $\wh\cK_{\df,3}^\uu$.

\section{Quantization condition with symmetric $\Kdf$}
\label{sec:2+1symm}

There are two drawbacks with the quantization condition (\ref{eq:QC3_2p1_asym}). First,
as it is derived using TOPT, $\wh\cK_{\df,3}^\uu$ is not Lorentz invariant.
This can be overcome, however,
 using an approach developed in BS2 that is based on Feynman diagrams and
leads to a quantization condition of exactly the same form, but with a Lorentz-invariant
version of the three-particle K matrix.
The second drawback is the asymmetric nature of $\wh\cK_{\df,3}^\uu$.
Different choices of the flavor indices lead to different
subclasses of contributions to three-to-three scattering, 
and the complete contribution is obtained only after summing them. 
This is a problem in practice because different parametrizations
are needed for each of the elements of $\wh\cK_{\df,3}^\uu$,
and many of the resulting parameters are redundant and do not affect the spectrum.

For this reason, a symmetric form of the quantization condition---one in which every
element of the three-particle K matrix includes all contributions---is highly desirable.
It was shown in BS2 how a symmetric quantization condition can be obtained
in the nondegenerate case using symmetrization identities. Furthermore, it was argued that
applying the symmetrization procedure to the result of a TOPT analysis 
yields a three-particle K matrix that is not only symmetric but also Lorentz invariant.
Here we apply this shortcut to obtain the symmetric and Lorentz-invariant
form of the quantization condition for the $2+1$ theory.

\subsection{Symmetrization identities}
\label{subsec:symmids}

These identities have been derived previously for three identical particles and
three nondegenerate particles in Refs.~\cite{\BSQC} and BS2, respectively.
The extension to the present case is tedious but straightforward, and we simply quote
the results. These involve rewriting $\wh F_G$ when it lies between two matrices both
with $\uu$ superscripts. Without loss of generality, we can replace the left-hand matrix
with a row vector $\boldsymbol X^{(u)}$ and the 
right-hand matrix with a column vector $\boldsymbol Z^{(u)}$,
since this corresponds to picking a particular element of the entire matrix product.
To be completely explicit, these vectors have the form
\begin{equation}
\boldsymbol X^{(u)} = \left( X^{(u,1)}_{k_1\ell m 1}, X^{(u,2)}_{k_2\ell m 2}\right)\,,
\qquad
\boldsymbol Z^{(u)} = \begin{pmatrix}
Z^{(u,1)}_{k_1\ell m 1} \\ Z^{(u,2)}_{k_2\ell m 2}
\end{pmatrix}\,,
\end{equation}
where the number in the superscripts indicates the spectator flavor, while the
$(u)$ indicates that $1\leftrightarrow 2$ symmetrization has yet to be performed.
It is important to keep in mind that, even after recombination with spherical
harmonics to obtain functions of $\{\bm k\}$, $X^{(u,1)}$ and $X^{(u,2)}$
are {\em different} quantities, involving sums over different diagrams.
The same holds for $Z^{(u,1)}$ and $Z^{(u,2)}$.

The identities can then be written in exactly the same form as in Ref.~\cite{\BSnondegen},
\begin{align}
\boldsymbol X^{(u)} \wh F_G \boldsymbol Z^{(u)}
&= 
\boldsymbol X^{(u)} \wh F \overrightarrow{\cS} \boldsymbol Z^{(u)} +
\boldsymbol X^{(u)} \wh{\overrightarrow{I}}_G \boldsymbol Z^{(u)}\,,
\label{eq:symmid1}
\\
&=
\boldsymbol X^{(u)} \overleftarrow{\cS} \wh F \boldsymbol Z^{(u)} +
\boldsymbol X^{(u)} \wh{\overleftarrow{I}}_G \boldsymbol Z^{(u)}\,,
\label{eq:symmid2}
\\
&=
\tfrac13\boldsymbol X^{(u)} \overleftarrow{\cS} \wh F \overrightarrow{\cS}\boldsymbol Z^{(u)} +
\boldsymbol X^{(u)} \wh{I}_{FG} \boldsymbol Z^{(u)}\,,
\label{eq:symmid3}
\end{align}
where $\overrightarrow{\cS}$ and $\overleftarrow{\cS}$ are symmetrization operators,
and the second terms in each identity involve matrices of integral operators.
We will not need the definitions of these operators, noting only that they sew together the
adjacent kernels into a combined infinite-volume object.

While having the same form, the identities do differ from those in BS2 in the details.
An obvious difference is that the matrices and vectors here are two-dimensional, whereas
those in the fully nondegenerate case are of dimension three.
A more subtle difference lies in the definition of the symmetrization operators.
Considering first the right-acting version, we have
\begin{equation}
\overrightarrow{\cS} \boldsymbol Z^{(u)}
=\begin{pmatrix} Z_{k_1\ell m 1} \\ Z_{k_2\ell m 2}/\sqrt2 \end{pmatrix}\,,
\label{eq:symmR}
\end{equation}
where both elements in the column vector are constructed from the
same underlying function $Z(\{\boldsymbol k\})$, but are written in $\{k_i\ell m i\}$
coordinates with different choices of spectator momentum.
The underlying function is
\begin{equation}
Z(\{\boldsymbol k\})=
Z^{(u,1)}(\{\boldsymbol k\}) + P_{11'} Z^{(u,1)}(\{\boldsymbol k\})
+ \sqrt{2} Z^{(u,2)}(\{\boldsymbol k\})\,,
\label{eq:Zsymm}
\end{equation}
where each of the components is reconstructed from the corresponding quantity with
$\{k_i\ell m i\}$ indices by combining with spherical harmonics.\footnote{%
In the notation of Ref.~\cite{\HSQCa}, $P_{11'} Z^{(u,1)} = Z^{(s,1)}$.}
Thus symmetrization adds the contributions from diagrams with all choices of spectator.
The factor of $\sqrt2$ in the third term in $Z$ arises from the $\sqrt2$ in the
off-diagonal elements of $\wh G$.
At first sight it appears to give extra weight to diagrams in which the third flavor is the
spectator, but this is canceled by the factor of $1/\sqrt2$ associated with the
second index that arises from the definitions of $\bra\alpha$ and $\ket\alpha$.

Finally, we note that a useful shorthand for the right-acting symmetrization operator
is given by
\begin{equation}
\overrightarrow{\cS} \boldsymbol Z^{(u)} = \ket\alpha\bra{\alpha_\cS} \boldsymbol Z^{(u)}\,,
\end{equation}
where $\ket\alpha$ and $\bra{\alpha_\cS}$ are defined above, and we have used the
fact that the kernels with spectators of flavor 2 are $1\leftrightarrow 1'$ symmetric.
In this shorthand, both the recombination with spherical harmonics,
and subsequent decomposition into $\{k_i\ell m i\}$ indices, are implicit.
Using this form, it is straightforward to check that
\begin{equation}
\overrightarrow{\cS} \overrightarrow{\cS}  = 3 \overrightarrow{\cS}\,,
\end{equation}
a result that holds also in the nondegenerate case, and which is used in the manipulations below.

The action of the left-acting symmetrization operator is simply expressed using this
shorthand. We find
\begin{equation}
\boldsymbol X^{(u)} \overleftarrow{\cS} 
=\boldsymbol X^{(u)} \ket{\alpha_\cS} \bra{\alpha} \,,
\end{equation}
from which it follows that
\begin{equation}
\overleftarrow{\cS} \overleftarrow{\cS}  = 3 \overleftarrow{\cS}\,.
\end{equation}

\subsection{Obtaining the symmetric quantization condition}
\label{subsec:QCsymm2p1}

With the symmetrization identities in hand, we now perform an arduous algebraic transformation
on $\wh \cM_{\df,3,L}^\uu$, replacing factors of $\wh F_G$ with $\wh F$s, while symmetrizing
the factors of $\wh \cK_{\df,3}^\uu$ and keeping track of the integral operators. 
The steps are algebraically identical to those in BS2
(see, in particular, appendix D of that work)
so we quote only the final results.
First, we note that, using Eqs.~(\ref{eq:M3L2p1final}) and (\ref{eq:M3Lhatdecomp}),
the finite-volume scattering amplitude can be decomposed as
\begin{equation}
\cM_{3,L,2+1} = \bra{\alpha_\cS} \wh \cM_{\df,3,L}^\uu \ket{\alpha_\cS} 
+ \bra{\alpha_\cS} \wh \cD_L^\uu \ket{\alpha_\cS}\,.
\label{eq:M3Ldecomp}
\end{equation}
The result of the algebraic manipulations is that
\begin{align}
\bra{\alpha_\cS} \wh \cM_{\df,3,L}^\uu \ket{\alpha_\cS} 
&=
\bra{\alpha_\cS} \wh \cM_{\df,3,L}^{\uu\prime\prime} \ket{\alpha_\cS} \,,
\\
\wh \cM_{\df,3,L}^{\uu\prime\prime}
&=
\left[\tfrac13 - \wh \cD_{23,L}^\uu  \wh F\right]
\wh \cK_{\df,3} \frac1{1+ \wh F_3 \wh\cK_{\df,3}}
\left[\tfrac13 - \wh F \wh \cD_{23,L}^\uu\right]
\label{eq:final1}
\end{align}
where
\begin{equation}
\wh F_3 = \frac{\wh F}3 - \wh F \frac1{\wh{\overline{\cK}}_{2,L}^{-1} + \wh F_G} \wh F\,,
\label{eq:F3hat2p1}
\end{equation}
while $\wh \cK_{\df,3}$ is a symmetrized three-particle K matrix
\begin{equation}
\wh \cK_{\df,3} = \overrightarrow{\cS} \wh \cK_{\df,3}' \overleftarrow{\cS}\,,
\label{eq:Kdf3hat2p1}
\end{equation}
with $\wh \cK_{\df,3}'$ related to $\wh\cK_{\df,3}^\uu$ by
\begin{align}
\wh \cK_{\df,3}' &= \wh Z \frac1{1 + \left[ \wh \cI_{FG} +
 \wh{\overleftarrow{\cI}}_G \wh{\overline{\cK}}_{2,L} \wh{\overrightarrow{\cI}}_G\right] \wh Z}\,,
\\
\wh Z &= \frac1{1+ \wh{\overline{\cK}}_{2,L} \wh{\overrightarrow{\cI}}_G} \wh\cK_{\df,3}^\uu
\frac1{1+ \wh{\overleftarrow{\cI}}_G \wh{\overline{\cK}}_{2,L}}\,.
\end{align}

We stress again that the meaning of $\wh \cK_{\df,3}$ being symmetric is
that all elements of the $2\times 2$ flavor matrix are obtained from a single underlying
quantity. The only subtlety concerns factors of $\sqrt 2$ arising from the definitions of
$\overrightarrow \cS$ and $\overleftarrow \cS$.
To make this explicit, we denote the underlying quantity by $\cK_{\df,3}$.
Then the matrix form is given by
\begin{equation}
\wh \cK_{\df,3} = \begin{pmatrix}
[\cK_{\df,3}]_{p_1 \ell' m' 1;k_1 \ell m 1} & [\cK_{\df,3}]_{p_1 \ell' m' 1;k_2 \ell m 2}/\sqrt2
\\
[\cK_{\df,3}]_{p_2 \ell' m' 2;k_1 \ell m 1}/\sqrt2 & [\cK_{\df,3}]_{p_2 \ell' m' 2;k_2 \ell m 2}/2
\end{pmatrix}
\,,
\label{eq:symmKdf3}
\end{equation}
where we are using the notation of Eq.~(\ref{eq:Kdf3tmat}).

From Eq.~(\ref{eq:final1}) we can read off the symmetrized form of the quantization condition,
namely
\begin{equation}
\det(1 + \wh F_3 \wh \cK_{\df,3}) = 0\,.
\label{eq:QC3pipiK}
\end{equation}
This has the same general form as all previous RFT quantization conditions involving symmetrized
kernels and only three-particle channels~\cite{\HSQCa,\BSQC,\BSnondegen}.
Indeed, it has identical form to the result
Eq.~(\ref{eq:QC3ndsym}) for the nondegenerate theory.
The difference, however, is that here the matrices $\wh F_3$ and $\wh \cK_{\df,3}$ are
two- rather than three-dimensional, and that here there are factors in 
the components of $\wh F_3$ that arise due to the presence of two identical
particles.

We summarize how to use the quantization condition  (\ref{eq:QC3pipiK}) in practice. 
One must choose a form for the underlying amplitude $\cK_{\df,3}$, for example the threshold
expansion described below in Sec.~\ref{sec:threshold}. This must be decomposed
into the $\{k_i\ell m i\}$ bases, and combined into the matrix form $\wh \cK_{\df,3}$ in
Eq.~(\ref{eq:symmKdf3}). The matrix $\wh F_3$, given in Eq.~(\ref{eq:F3hat2p1}),
is constructed from $\wh F$, $\wh F_G$ and $\wh{\overline{\cK}}_{2,L}$,
which are given in Eqs.~(\ref{eq:Fhat2p1}), (\ref{eq:FG2p1}) and (\ref{eq:K2Lhat2p1}),
respectively. Finally, one must truncate the sums over $\ell$, leading to a finite matrix problem

We can also obtain the integral equations relating $\cM_3$ to $\cK_{\df,3}$ from the above
results. This is achieved by taking the $L\to\infty$ limit of the result for $\cM_{3,L}$, 
contained in Eqs.~(\ref{eq:M3Ldecomp}), (\ref{eq:final1}), and (\ref{eq:DLhat}),
in the manner explained in Ref.~\cite{\HSQCb} and outlined in the previous section.
We do not write out the integral equations explicitly, 
as they can be obtained by a simple generalization
of the results given explicitly in Refs.~\cite{\BSQC} and BS2.
This is because the key equations (\ref{eq:M3Ldecomp}), (\ref{eq:final1}), and (\ref{eq:DLhat})
have essentially identical form to those for the nondegenerate theory
given in Eqs.~(\ref{eq:M3hatres}), (\ref{eq:symMdf3L}), and (\ref{eq:DL}), respectively.
An important consequence of these integral equations is that they can be used to show
that $ \cK_{\df,3}$ is Lorentz invariant, under the assumption that the relations between
$\cM_3$ and $\wh \cK_{\df,3}$ are invertible.

We show in Appendix~\ref{sec:2d+1}
that the results of this subsection can be obtained by taking an appropriate
projection of those for two degenerate but nonidentical particles combined with a third.

We stress that the complicated nature of the relation between the asymmetric kernel
$\wh\cK_{\df,3}^\uu$ and the symmetrized version $\wh \cK_{\df,3}$ has no practical import.
In a generic effective field theory we do not have knowledge of either of these quantities, 
and must parametrize them prior to use in their respective quantization conditions.
From this viewpoint, the symmetrized quantization condition is preferable as it is
described by fewer parameters.

\section{Threshold expansion of $\wh{\cK}_{\df,3}$}
\label{sec:threshold}

A numerical implementation of the quantization condition for the $2+1$ theory
requires a parametrization of the three-particle K matrix, which we stress
is an infinite-volume quantity. Here we focus on the underlying quantity that appears
in the symmetric version, i.e.~on $\cK_{\df,3}$.
We present its parametrization in an expansion about threshold,
generalizing the systematic method introduced for the degenerate case in Ref.~\cite{\dwave}.
We first describe the Lorentz invariants that are required, 
then give the result for the nondegenerate K matrix, 
for which there are less symmetries and for which the expansion was not given in BS2,
and finally describe the result for the $2+1$ theory.

As explained in the previous section, although $\wh{\cK}_{\df,3}$ is a matrix, each element contains
the same underlying quantity, which is a function of the three incoming and 
three outgoing on-shell momenta. 
It is the latter quantity that we parametrize here. 
To use the resulting form in the quantization condition it must be decomposed
into $\{k_i \ell m i\}$ coordinates for each choice of spectator flavor $i$.
This is a straightforward exercise that was discussed for the case of three identical particles in 
Ref.~\cite{\dwave}. The generalizations to the $2+1$ and nondegenerate theories are
unilluminating, so we do not present them explicitly.

The corresponding underlying symmetric K matrix
in the nondegenerate theory was called $\wt \cK_{\df,3}$ in BS2, 
but here, following Appendix~\ref{sec:recap}, we use $\wt \cK_{\df,3;\nd}$
[see Eq.~(\ref{eq:Kdf3tmat})].

The properties of $\Kdf$ and $\wt{\cK}_{\df,3;\nd}$ that we use are that they are real 
(due to our use of the principle value prescription in integrals), 
Lorentz invariant, invariant under time reversal and parity symmetries
(assuming these discrete symmetries hold in the underlying theory),
and, in the case of the $2+1$ theory, invariant under interchange of the two identical
particles, separately for the initial and final states.

\subsection{Lorentz invariants for three-particle scattering}
\label{subsec:Mandelstams}

We denote the incoming (outgoing) on-shell 4-momenta as $p_i$ ($p'_i$)
with flavors here taken to be $i\in\{1,2,3\}$.
The following discussion holds both for the fully nondegenerate and $2+1$ theories,
with only the masses differing in the two cases.
We use the generalized Mandelstam variables,
\begin{equation}
    s \equiv  E^{*2} \,, \ \
    s_{i} \equiv (p_j+p_k)^2, \ \ s_{i}' \equiv (p'_j+p'_k)^2\,, \ \ 
    t_{ij} \equiv (p_i-p_j')^2\,,
    \label{eq:Mandelstams}
\end{equation}
where $i$, $j$, and $k$ are ordered cyclically.
It is convenient to use dimensionless quantities which vanish at threshold,
\begin{equation}
    \Delta \equiv \frac{s - M_\Sigma^2}{M_\Sigma^2}\,, \ \
    \Delta_i \equiv \frac{s_{i} - (m_j+m_k)^2}{M_\Sigma^2} \,, \ \
     \Delta_i' \equiv \frac{s_{i}' - (m_j+m_k)^2}{M_\Sigma^2}\,, \ \
    \widetilde{t}_{ij} \equiv \frac{t_{ij} - (m_i-m_j)^2}{M_\Sigma^2}\,,
    \label{eq:Mandelstams_reduced}
\end{equation}
where
\begin{align}
	M_\Sigma\equiv m_1+m_2+m_3 \,.
\end{align}
The sixteen quantities in Eq.~\eqref{eq:Mandelstams_reduced} 
are constrained by eight independent relations,
\begin{align}
    \sum_{i=1}^3\Delta_{i} &= \sum_{i=1}^3\Delta_{i}' = \Delta   \label{eq:constraint_1} \\
    \sum_{j=1}^3 \widetilde{t}_{ij} &= \Delta_{i} - \Delta, \quad \sum_{j=1}^3 \widetilde{t}_{ji} = \Delta_{i}' - \Delta \qquad \forall i\in\{1,2,3\}\,.
     \label{eq:constraint_2}
\end{align}
Thus only eight are independent: 
the overall CM energy (parametrized here by $\Delta$)
and seven ``angular'' degrees of freedom.\footnote{%
We call these variables angular since they span a compact space.}
This counting is as expected: six on-shell momenta 
with total incoming and outgoing 4-momentum fixed have $3\cdot6-4\cdot2=10$ degrees of freedom, which is reduced to 7 by overall rotation invariance.

For physical scattering,
it is straightforward to show that $\Delta_{i}$, $\Delta_{i}'$, and $-\widetilde{t}_{ij}$ 
are all non-negative, and the constraint equations then lead to the inequality
\begin{equation}
    0 \leq \Delta_{i}, \Delta_{i}', -\widetilde{t}_{ij} \leq \Delta\,. 
    \label{eq:thresh_scale}
\end{equation}
Thus all the variables
$\{ \Delta, \Delta_{i}, \Delta_{i}', \widetilde{t}_{ij} \}$ 
can be treated as being of the same order in an expansion about threshold.

\subsection{Threshold expansion for nondegenerate theory}
\label{subsec:K3_expansion_nd}

By construction, $\wt{\cK}_{\df,3}$ is a smooth function for some region around threshold.\footnote{%
There can be dynamical singularities due to three-particle resonances, but, generally,
these will lie away from threshold.}
Thus it can be expanded in a Taylor series in the variables
$\{ \Delta, \Delta_{i}, \Delta_{i}', \widetilde{t}_{ij} \}$, which are all treated as being
of $\cO(\Delta)$.
Since $\Kdf$ is real, the coefficients in this expansion must also be real.
The expansion must also respect T and P symmetries, which here implies symmetry
under the simultaneous interchange $p_i \leftrightarrow p'_i$ for all $i$.
In terms of our variables this implies that $\Kdf$ must be unchanged when
$\Delta_{i} \leftrightarrow \Delta_{i}' $ and, simultaneously,
$  \widetilde{t}_{ij}\leftrightarrow \widetilde{t}_{ji}$, for all $i,j$.
It is then a tedious but straightforward exercise to write down the allowed terms
at each order in $\Delta$, and reduce them to an independent set using
the constraints \eqref{eq:constraint_1} and \eqref{eq:constraint_2}.

At leading (zeroth) order, we simply have 
$\wt\cK_{\df,3;\nd} = \wt \cK_{\df,3;\nd}^{\text{iso},0} = \text{constant}$.
At linear order in $\Delta$, there are 10 terms that respect TP symmetry,
\begin{itemize}
	\item $\Delta$
	\item $\Delta_i^S \equiv \Delta_i + \Delta_i'$ for $i\in\{1,2,3\}$
	\item $\wt t_{ij}^S \equiv \wt t_{ij} + \wt t_{ji} = \wt t_{ji}^S$ for $1\leq i \leq  j \leq 3$.
\end{itemize}
The constraints imply four redundancies
\begin{equation}
	\sum_{i=1}^3 \Delta_i^S = 2\Delta\,,
	\qquad
	\sum_{j=1}^3 \wt t_{ij}^S = \Delta_i^S - 2\Delta \qquad \forall i\in\{1,2,3\} \,,
\end{equation}
and thus six independent first-order terms.
A possible parametrization (chosen with an eye towards the $2+1$ case) is
\begin{multline}
	\wt \cK_{\df,3;\nd} = \wt \cK_{\df,3;\nd}^{\text{iso},0}
	+ \wt \cK_{\df,3;\nd}^{\text{iso},1} \Delta
	+ \wt \cK_{\df,3;\nd}^{A,1} (\Delta^S_1 - \Delta^S_2)
	+ \wt \cK_{\df,3;\nd}^{B,1} \Delta^S_3 
	\\
	+ \wt \cK_{\df,3;\nd}^{C,1} (\wt t^S_{11}+\wt t^S_{22} - 2 \wt t^S_{12}) 
	+ \wt \cK_{\df,3;\nd}^{D,1} (\wt t^S_{13} - \wt t^S_{23})
	+ \wt \cK_{\df,3;\nd}^{E,1} \wt t^S_{33}
	+ \mc{O}(\Delta^2) \,,
	\label{eq:Kdfparam_nd}
\end{multline}
where $\{\cK_{\df,3;\nd}^{\text{iso},0}, \cK_{\df,3;\nd}^{\text{iso},1}, \cK_{\df,3;\nd}^{A,1}, 
\cK_{\df,3;\nd}^{B,1}, \cK_{\df,3;\nd}^{C,1}, \cK_{\df,3;\nd}^{D,1}, \cK_{\df,3;\nd}^{E,1}\}$ 
are parameters that---along with those in the effective-range expansion of 
$\cK_2$---could, 
in principle, be obtained from a fit to finite-volume spectra from lattice calculations. 
This should be compared to a single parameter (that proportional to $\Delta$)
that is present at linear order for three identical particles~\cite{\dwave}.

Although we do not explicitly work out the $\{k_i \ell m i\}$ decomposition, we do note that,
while the isotropic terms involve only $\ell=\ell' =0$ contributions,
the full set of $\cO(\Delta)$ terms leads to generic p-wave contributions,
i.e.~with $\{\ell=1$, $\ell'=0\}$, $\{\ell=0$, $\ell'=1\}$, and $\{\ell=1$, $\ell'=1\}$.
Such terms are allowed because the three particles are distinguishable.

\subsection{Threshold expansion for $2+1$ theory}
\label{subsec:K3_expansion_2p1}

We now turn to the the $2+1$ theory. In this subsection,
we will continue to use the nondegenerate flavor labels $\{1,2,3\}$, 
rather than the $\{1,1',2\}$ used in previous sections,
in order to reduce the number of primes.

The additional symmetries in the $2+1$ theory are those under $1\leftrightarrow 2$ interchange
separately for the initial and final particles. This considerably reduces the number of independent
terms, since three of the linear terms in Eq.~(\ref{eq:Kdfparam_nd}) are odd under some combination of the interchange transformations. Thus at linear order there are three parameters,
\begin{equation}
	\cK_{\df,3} =  \cK_{\df,3}^{\text{iso},0}
	+ \cK_{\df,3}^{\text{iso},1} \Delta
	+ \cK_{\df,3}^{B,1} \Delta^S_3 
	+ \cK_{\df,3}^{E,1} \wt t^S_{33}
	+ \mc{O}(\Delta^2) \,.
	\label{eq:Kdfparam_2p1}
\end{equation}
When this form is decomposed into $\{k_i\ell m i\}$ indices, only s-wave terms occur if
the spectator is chosen as flavor 3 (consistent with the absence of odd partial waves in the
interaction of the two identical particles), but there will be both s- and p-wave terms if the spectator has
flavor 1 or 2.
These results are consistent with one another because, even when the spectator has flavor 3,
there can be nonzero angular momentum between the interacting pair and the spectator.

Given the relatively small number of additional parameters at linear order in $\Delta$, we have worked
out the independent terms at quadratic order. We find nine in all. One choice of basis for the terms
is
\begin{multline}
\big\{
\Delta^2, \ \Delta\Delta_3^S,\ (\Delta_3^S)^2,\ 
\Delta \wt t_{33}^S,\ \Delta_3^S \wt t_{33}^S,\ (\wt t_{33}^S)^2,\
\Delta_3 \Delta'_3,\ \wt t_{11} \wt t_{22} + \wt t_{12} \wt t_{21},\
(\wt t_{11})^2+ (\wt t_{22})^2 +(\wt t_{12})^2 + (\wt t_{21})^2 \big\}\,,
\end{multline}
where the first six terms are simply the products of the linear terms, while the last three are
additional forms. Note that the last three do not involve the superscript $S$.

\section{Conclusions}
\label{sec:conc}

We have presented in this work the  formalism needed to extract three-particle
scattering amplitudes from finite-volume energy levels in a system with
two identical particles together with a third that is different.
The formalism allows for relativistic particles, and does not make any truncation
of angular momentum indices.
It extends previous work in the RFT approach~\cite{\HSQCa,\HSQCb,\BHSQC,\BHSK,\isospin,\BSQC,\BSnondegen}.

Our main results are as follows. First, the asymmetric form of the quantization condition
is given by Eq.~(\ref{eq:QC3_2p1_asym}), 
which contains the three-particle K matrix $\wh \cK_{\df,3}^\uu$,
itself related to the three-particle scattering amplitude in the $2+1$ theory, $\cM_{3,2+1}$,
by the integral equations given implicitly by Eqs.~(\ref{eq:M3L2p1final})-(\ref{eq:D23Lhat}).
Second, the symmetric form of the quantization condition is given by
Eq.~(\ref{eq:QC3pipiK}), and contains the symmetric K matrix $\wh \cK_{\df,3}$, which is
related to $\cM_{3,2+1}$ by integral equations summarized in Sec.~\ref{subsec:QCsymm2p1}.
Both forms of the quantization condition involve two-dimensional matrices,
in contrast to the three-dimensional matrices needed for three nondegenerate particles.
The quantities entering these quantization conditions are built from the kinematic
matrices $\wh F$ and $\wh G$, which are given in Eqs.~(\ref{eq:Fhat2p1}) and
(\ref{eq:Ghat2p1}), respectively, and a matrix form of the two-particle K matrix,
$\wh{\overline{\cK}}_{2,L}$, given in Eq.~(\ref{eq:K2Lhat2p1}). 
The symmetry factors associated with the presence of two identical particles are
contained in $\wh G$ and $\wh{\overline{\cK}}_{2,L}$.

This extension of the three-particle formalism allows its application 
to several new systems in QCD.
Those which are most likely to be studied in the near term are 
$\pi^+\pi^+ K^+$ and $\pi^+ K^+ K^+$ (or their charge conjugates),
since these channels do not involve diagrams with quark-antiquark annihilation.
To aid in the application of the formalism, we have provided the first two terms in the
threshold expansion for the three-particle K matrices that enter at intermediate stages.
The numerical implementation of the formalism should not be substantially more challenging
than that used in previous successful applications for the
$3\pi^\pm$ and $3K^\pm$ systems~\cite{\ThreeQCDNumerics}.

As noted in the introduction, the NREFT formalism has been generalized to a system
with two degenerate particles and a third in Ref.~\cite{\DDK}, keeping only s-wave interactions.
The specific application considered in that work is the $I=1/2$ $DDK$ system,
which contains two channels ($D^+ D^0 K^+$ and $D^+ D^+ K^0$ for $I_z=1/2$),
and thus lies beyond the scope of the present work.\footnote{%
In fact, strictly speaking the analysis of Ref.~\cite{\DDK} is incomplete, as the
$D^0 D_s^+ \pi^0$ and $D^+ D_s^+ \pi^-$ channels (which open at lower CMF energy)
should also be included, although the coupling to these channels may be weak.}
Nevertheless, we can make a broad-brush comparison with the formalism of Ref.~\cite{\DDK}.
In particular, we note that both approaches use two-dimensional flavor matrices,
and that the $\wh G$ matrices have similar forms (with a zero in the bottom right entry).

We have been able to shorten the derivation 
by reusing many of the algebraic manipulations used in previous work.
In addition, we have recast the result for three identical particles in a form that
makes the symmetry factors explicit, allowing a simple generalization to the $2+1$ theory.
We stress that, although both asymmetric and symmetric forms of the quantization conditions
have the same general appearance as that for three nondegenerate particles, this superficial equivalence
hides important differences. In particular, the flavor matrices here are two-
rather than three-dimensional, and they contain embedded symmetry factors.

Looking forward, to obtain a completely general three-particle formalism 
requires allowing for multiple three-particle channels that are nondegenerate, 
including particles with spin, and
applying these generalizations to cases with two- and three-particle channels.
Work on these extensions is underway.

\section*{Acknowledgments}

This work was supported in part 
by the U.S. Department of Energy award no. DE-SC0011637.

\appendix

\section{Summary of results for three nondegenerate particles}
\label{sec:recap}

We recall here the results and notation from BS2 that are needed in the main text.
These concern a theory
containing three real scalar fields, $\phi_i$, with physical masses $m_i$,
where $i=1-3$ labels the flavor.
The results in BS2
concern the three-particle state containing one particle of each flavor.
We note that we have added superscripts or subscripts ``$\nd$'' to the quantities
appearing in this appendix in order to distinguish them from the corresponding
quantities in the $2+1$ theory. These superscripts and subscripts are absent in BS2.

We begin by explaining the matrix indices that are used in all the results,
which are $p \ell m i$.
These describe a configuration of the on-shell three-particle state.
Here $p \ell m$ are the standard indices used in the RFT approach,
which were introduced in Ref.~\cite{\HSQCa}, while the additional flavor index $i$ is
needed for distinguishable particles.
We begin by noting that
an on-shell state is described by a triplet of three-momenta 
$\{{\bm p}_1,{\bm p}_2,{\bm p}_3\}\equiv \{\bm p\}$ (with the subscript indicating the flavor)
that satisfy
\begin{equation}
\bm p_1+\bm p_2 + \bm p_3 = \bm P\ \ {\rm and} \ \
\omega^{(1)}_{p_1} + \omega^{(2)}_{p_2} + \omega^{(3)}_{p_3} = E\,.
\label{eq:momcons}
\end{equation}
Here the superscript on $\omega$ indicates the mass used to calculate the on-shell energy,
\begin{equation}
\omega^{(i)}_{p} = \sqrt{\bm p^2 + m_i^2}\,.
\label{eq:omegai}
\end{equation}
We divide the three particles into a ``spectator'' and a ``nonspectator pair.''
There are three ways to do this, corresponding to the choice of flavor of the spectator,
and this choice is denoted $i$ in the index set.
The spectator momentum is then drawn from the finite-volume set,
and denoted $\bm p$, or simply $p$ in the index set.
The other two particles are boosted to their CMF,
with the resulting momenta being ${\bm p}_j^{*(i,j,p_i)}$ and ${\bm p}_k^{*(i,k,p_i)}$, 
where the flavors $i$, $j$, and $k$ are taken to lie in cyclic order.
The notation in the superscript is defined as follows. The $*$ indicates a boosted momentum,
and, in $(i,j,p_i)$, the first entry indicates the flavor of the spectator particle, the second the flavor
of the particle that is being boosted, and the third the momentum of the spectator particle.\footnote{%
This is a more explicit notation than used in BS2, where the subscript on the momenta
played double duty as a flavor label. Here that is not always possible.
We also note that the same definition of the boosted momenta can be used if one of
the three particles is not on shell, as is the case in $\wt F^{(i)}$ and $\wt G^{(ij)}$ below.}
The magnitudes of the boosted momenta are fixed by momentum conservation to be
\begin{align}
\left|{\bm p}_j^{*(i,j,p_i)}\right|^2 = \left|{\bm p}_k^{*(i,k,p_i)}\right|^2 =
 \big(q_{2,p_i}^{*(i)}\big)^2  &\equiv \frac{\lambda(\sigma_i,m_j^2, m_k^2)}{4\sigma_i}\,, \quad
\sigma_i \equiv (E-\omega^{(i)}_{p_i})^2 - (\bm P-\bm p_i)^2\,,
\end{align}
where $\lambda(a,b,c) = a^2+b^2+c^2 -2 ab -2 ac - 2 bc$ is the standard triangle function.
This angular dependence is then decomposed into
spherical harmonics relative to the direction $\wh{\bm p}_j^{\;*(i,j,p_i)}$,
leading to the $\ell m$ indices.
An on-shell quantity $X(\{\bm p\})$ can be reconstructed using
\begin{equation}
X(\{\bm p\}) = \sum_{\ell m} X_{p_i\ell m i} \sqrt{4\pi} Y_{\ell m}(\wh{\bm p}_j^{*(i,j,p_i)})\,.
\label{eq:Xdecomp}
\end{equation}
An important point in the following is that, 
had we used $\wh{\bm p}_k^{\;*(i,k,p_i)}$ to define the spherical harmonics rather
than $\wh{\bm p}_j^{\;*(i,j,p_i)}$,
the resulting $X_{p_i\ell m i}$ would have differed by a factor of $(-1)^\ell$.

There are three different ways of decomposing a given quantity $X$,
one for each choice of flavor index $i$.
The quantization condition makes use of this freedom.
For example, the three-particle interaction kernel\footnote{%
The tilde on $\wt \cK_{\df,3;\nd}$ is a notation inherited from BS2,
which we continue to use for consistency. The subscript ``nd" is absent in BS2 but
added here to distinguish this quantity from the similar one for the $2+1$ theory
that appears in the main text.}
 ${\wt{\cK}}_{\df,3;\nd}(\{\bm p\};\{\bm k\})$,
which appears in the symmetric form of the quantization condition described below,
is decomposed in all nine possible ways---three choices for the incoming momenta combined with three for the outgoing momenta.
These are collected into a $3\times3$ flavor matrix
\begin{equation}
\wh{\wt{\cK}}_{\df,3;\nd} = \begin{pmatrix}
[\wt\cK_{\df,3;\nd}]_{p_1\ell' m' 1;k_1 \ell m 1} & 
[\wt\cK_{\df,3;\nd}]_{p_1 \ell' m' 1;k_2 \ell m 2} & 
[\wt\cK_{\df,3;\nd}]_{p_1 \ell' m' 1;k_3 \ell m 3}
\\
[\wt\cK_{\df,3;\nd}]_{p_2 \ell' m' 2;k_1 \ell m 1} & 
[\wt\cK_{\df,3;\nd}]_{p_2 \ell' m' 2;k_2 \ell m 2}  & 
[\wt\cK_{\df,3;\nd}]_{p_2 \ell' m' 2;k_3 \ell m 3}
\\
[\wt\cK_{\df,3;\nd}]_{p_3\ell' m' 3;k_1 \ell m 1} & 
[\wt\cK_{\df,3;\nd}]_{p_3\ell' m' 3;k_2 \ell m 2} & 
[\wt\cK_{\df,3;\nd}]_{p_3\ell' m' 3;k_3 \ell m 3}
\end{pmatrix}
\,.
\label{eq:Kdf3tmat}
\end{equation}
The caret (or ``hat'') on $\wh{\wt{\cK}}_{\df,3;\nd}$ is the notation we use throughout
to indicate a matrix in flavor space.
We have added a flavor label on the $p_i$ and $k_i$ in order to emphasize which
momenta from the initial and final sets are picked out as the spectators,
although this is a redundant choice in the presence of the flavor index.

As a second example of the matrix notation, we consider
the three-particle kernel $\cK_{\df,3;\nd;i,j}(\{\bm p\};\{\bm k\})$
that  appears in the asymmetric form of quantization condition to be discussed shortly. Here the kernel itself has flavor indices,
indicating that it depends upon the choice of spectators. The decomposition into
$\{p\ell m i\}$ indices is done accordingly, leading to
\begin{equation}
\wh{{\cK}}_{\df,3;\nd} = \begin{pmatrix}
[\cK_{\df,3;\nd;1,1}]_{p_1\ell' m' 1;k_1 \ell m 1} & 
[\cK_{\df,3;\nd;1,2}]_{p_1 \ell' m' 1;k_2 \ell m 2} & 
[\cK_{\df,3;\nd;1,3}]_{p_1 \ell' m' 1;k_3 \ell m 3}
\\
[\cK_{\df,3;\nd;2,1}]_{p_2 \ell' m' 2;k_1 \ell m 1} & 
[\cK_{\df,3;\nd;2,2}]_{p_2 \ell' m' 2;k_2 \ell m 2}  & 
[\cK_{\df,3;\nd;2,3}]_{p_2 \ell' m' 2;k_3 \ell m 3}
\\
[\cK_{\df,3;\nd;3,1}]_{p_3\ell' m' 3;k_1 \ell m 1} & 
[\cK_{\df,3;\nd;3,2}]_{p_3\ell' m' 3;k_2 \ell m 2} & 
[\cK_{\df,3;\nd;3,3}]_{p_3\ell' m' 3;k_3 \ell m 3}
\end{pmatrix}
\,.
\label{eq:Kdf3mat}
\end{equation}
We stress that the difference between this matrix and that in Eq.~(\ref{eq:Kdf3tmat})
is that here each entry is intrinsically different, in addition to being decomposed into
different coordinates. For this reason we refer to $\wh\cK_{\df,3;\nd}$ as an asymmetric kernel,
in contrast to $\wh{\wt{\cK}}_{\df,3;\nd}$, which we refer to as a symmetric kernel.

We mostly focus in this paper on two of the three forms of the quantization condition
derived in BS2. The first, derived using TOPT, involves the asymmetric three-particle
kernel of Eq.~(\ref{eq:Kdf3mat}), and takes the form
\begin{equation}
\det\left(1 + \wh F_G^\nd [\wh{\overline{\cK}}_{2,L} + \wh \cK_{\df,3;\nd}] \right) = 0\,,
\label{eq:QC3ndasym}
\end{equation}
where the determinant runs over all four matrix indices.
Infinite-volume two-particle K matrices are contained in
\begin{align}
\wh{\overline{\cK}}_{2,L;\nd} &= {\rm diag}
\left( \overline{\cK}_{2,L}^{(1)},  \overline{\cK}_{2,L}^{(2)},  \overline{\cK}_{2,L}^{(3)}\right)\,,
\label{eq:K2Lhat}
\\
\left[\overline{\cK}_{2,L}^{(i)}\right]_{p_i\ell' m'; r_i \ell m} &= 
\delta_{\bm p_i \bm r_i} 2\omega^{(i)}_{r_i} L^3 \cK^{(i)}_{2;\ell' m';\ell m}(\bm r_i)\,,
\label{eq:K2Li}
\\
\cK^{(i)}_{2;\ell' m';\ell m}(\bm r_i)
&=
\delta_{\ell' \ell} \delta_{m' m}  \cK_{2,\ell}^{(i)}(q_{2,r_i}^{*(i)})\,,
\label{eq:K2defend}
\end{align}
with $\cK_{2,\ell}^{(i)}$ being the $\ell$th partial wave of the infinite-volume two-particle
K matrix for scattering of flavors $j$ and $k$. Here $i$, $j$, and $k$ are in cyclic order, 
and the partial-wave decomposition is defined relative to the direction of the momentum of
the particle of flavor $j$ in the pair CMF:
\begin{equation}
\cK_2^{(i)}(\bm p_j,\bm p_k;\bm r_j,\bm r_k) =   \sum_{\ell' m' \ell m}
\cK^{(i)}_{2;\ell' m';\ell m}(\bm p_i) 
4\pi  Y_{\ell' m'}(\bm p_j^{*(i,j,r_i)}) Y_{\ell m}(\bm r_j^{*(i,j,r_i)})\,.
\end{equation}
For completeness
we give the relation between the s-wave K matrix and the corresponding phase shift
and scattering length above threshold,
\begin{equation}
\frac1{\cK_{2,\ell=0}^{(i)}(q^*)} = \frac{q^*}{8\pi \sqrt{\sigma_i}} \cot \delta_0(q^*)
= \frac{1}{8\pi \sqrt{\sigma_i}} \left[ -\frac{1}{a_0} + \cO(q^{*2}) \right]
\,.
\label{eq:Ktodelta}
\end{equation}
The normalization here is that for distinguishable particles.
When we consider the $2+1$ system in the main text, we encounter the two-particle K matrix for identical particles,
i.e.~$\overline{\cK}_{2,L}^{(2)}$ in
Eq.~(\ref{eq:K2Lhat2p1}). In this case, the relation to $\cot\delta_0$ contains
an extra overall symmetry factor of $1/2$, i.e.~the $8\pi$ in Eq.~(\ref{eq:Ktodelta}) is
changed to $16\pi$.

The matrix $\wh F_G^\nd$ contains finite-volume kinematic factors:
\begin{align}
\wh F_G^\nd &= \wh F^\nd+ \wh G^\nd\,,
\label{eq:FGhat}
\\
\wh F^\nd &= {\rm diag}
\left( \wt F^{(1)},\wt F^{(2)},\wt F^{(3)} \right)\,,
\label{eq:Fhat}
\\
\wh G^\nd &=
\begin{pmatrix}
0 & \wt G^{(12)} P_L  & P_L\wt G^{(13)} \\
P_L \wt G^{(21)} & 0 & \wt G^{(23)} P_L \\
\wt G^{(31)} P_L & P_L \wt G^{(32)} & 0
\end{pmatrix}\,,
\label{eq:Ghat}
\end{align}
the entries of which are defined in the following.

The second form of the quantization condition involves the symmetric kernel of Eq.~(\ref{eq:Kdf3tmat}),
 and is
 \begin{equation}
 \det\left(1 + \wh F_3^\nd \wh{\wt{\cK}}_{\df,3;\nd} \right) = 0\,,
 \label{eq:QC3ndsym}
 \end{equation}
where
\begin{align}
\wh F_3^\nd &= \tfrac13 \wh F^\nd - \wh F^\nd \wh \cD_{23,L}^\nd \wh F^\nd \,, 
\label{eq:F3hat}
\\
\wh \cD_{23,L}^\nd
&=  \wh{\overline{\cM}}{}_{2,L}^\nd
\frac1{1 +\wh G^\nd  \wh{\overline{\cM}}{}_{2,L}^\nd }\,.
\label{eq:D23L}
\\
(\wh{\overline{\cM}}{}_{2,L}^\nd)^{-1} &= (\wh{\overline{\cK}}{}_{2,L}^\nd)^{-1} + \wh F^\nd\,.
\label{eq:M2Lhat}
\end{align}

The kinematic functions contained in $\wh F_G^\nd$ are defined by
\begin{align}
\left[\wt F^{(i)}\right]_{p' \ell' m';p \ell m} =
\delta_{\bm p' \bm p} \frac{H^{(i)}(\bm p)}{2\omega_{p}^{(i)} L^3}
\left[ \frac1{L^3} \sum_{\bm r}^{\rm UV} - \PV \int^{\rm UV} \frac{d^3 r}{(2\pi)^3} \right]
\Bigg[
\frac{\cY_{\ell' m'}(\bm r^{*(i,j,p)})}{\big(q_{2,p}^{*(i)}\big)^{\ell'}}
& \frac1{4\omega_{r}^{(j)} \omega_{b}^{(k)} \big(E\!-\!\omega_{p}^{(i)}\!-\!\omega_{r}^{(j)}\!-\!\omega_{b}^{(k)}\big)}
\label{eq:Ft}
\\
& \hspace{64pt} \times
\frac{\cY_{\ell m}(\bm r^{*(i,j,p)})}{\big(q_{2,p}^{*(i)}\big)^{\ell}}
\Bigg] \nonumber
\,,
\end{align}
with $i$, $j$, and $k$ in cyclic order, $\bm b= \bm P-\bm p-\bm r$, and
\begin{equation}
\left[\wt G^{(ij)}\right]_{p \ell' m';r \ell m} = 
\frac1{2\omega^{(i)}_{p} L^3}
\frac{\cY_{\ell' m'}(\bm r^{*(i,j,p)})}{\big(q_{2,p}^{*(i)}\big)^{\ell'}}
\frac{H^{(i)}(\bm p) H^{(j)}(\bm r)}{b_{ij}^2-m_k^2}
\frac{\cY_{\ell m}(\bm p^{*(j,i,r)})}{\big(q_{2,r}^{*(j)}\big)^{\ell}}
\frac1{2\omega^{(j)}_{r} L^3}\,,
\label{eq:Gt}
\end{equation}
where the four-vector $b_{ij}$ given by
\begin{equation}
b_{ij} = (E-\omega_{p}^{(i)}-\omega_{r}^{(j)},\bm P-\bm p-\bm r)\,.
\end{equation}
One choice for the cutoff functions $H^{(i)}(\bm p)$ is~\cite{\BSnondegen},
\begin{align}
H^{(i)}(\bm p_i) &= J(z_i)\,,\quad
z_i = (1+\epsilon_H) \frac{\sigma_i}{(m_j+m_k)^2}\,,
\\
J(z) &= \left\{ 
\begin{array}{ll} 0, & z \le 0 \\ 
\exp\left(-\frac1z\exp\left[-\frac1{1-z}\right]\right), & 0 < z < 1\\
1, & 1 \le z \,, \end{array}\right.
\end{align}
with $\epsilon_H$ being a positive constant for which a practical value is $\approx 0.1$.
We note that this form is Lorentz invariant.
The harmonic polynomials are 
\begin{equation}
\cY_{\ell m}(\bm{a})
= \sqrt{4\pi} Y_{\ell m}(\widehat{\bm a}) |\bm a|^\ell\,,
\label{eq:harmonicpoly}
\end{equation}
with the spherical harmonics chosen to be in the real basis.
The boost used to define arguments of the harmonic polynomial is that from Ref.~\cite{\HSQCa}.
Finally, the parity operators in $\wh G$ are 
\begin{equation}
\left[P_L\right]_{p\ell'm';k\ell m} = \delta_{\bm p \bm k} \delta_{\ell' \ell} \delta_{m' m} 
(-1)^\ell \,,
\label{eq:PLdef}
\end{equation}
and arise because of a mismatch between the cyclic convention used to define spherical
harmonics for kernels and the convention used in the definition of $\wt G^{(ij)}$.

We now describe the relation between the (infinite-volume) scattering amplitude $\cM_{3;\nd}$ 
and the symmetric kernel $\wt\cK_{\df,3;\nd}$. The analogous relation for the asymmetric
kernel $\cK_{\df,3;\nd}$ is given in BS2, and not needed here.
The relation involves integral equations, and is given implicitly by
\begin{align}
\wh \cM_{3;\nd} &= \lim_{L\to \infty} \wh \cM_{3,L;\nd} \,,
\label{eq:M3hatres}
\\
\wh \cM_{3,L;\nd} &=
\overrightarrow \cS_{\rm nd} \left(
\wh \cM''_{\df,3,L;\nd} + \wh \cD_L^\nd
\right) \overleftarrow\cS_{\rm nd}\,,
\label{eq:M3hatresb}
\end{align}
where $\overrightarrow \cS_{\rm nd}$ and $ \overleftarrow \cS_{\rm nd}$ are symmetrization operators
defined in BS2,
and the infinite-volume limit must be taken with the $i\epsilon$ prescription applied to
poles~\cite{\HSQCb}.
The quantities entering Eq.~(\ref{eq:M3hatresb}) are
\begin{equation}
\wh \cM''_{\df,3,L;\nd} 
=
\left[\tfrac13 - \wh \cD_{23,L}^\nd  \wh F^\nd \right] 
\wh{\wt{\cK}}_{\df,3;\nd}
\frac1{1+ \wh F_3^\nd \wh{\wt{\cK}}_{\df,3;\nd} }
\left[\tfrac13 - \wh F^\nd  \wh \cD_{23,L}^\nd \right] 
\label{eq:symMdf3L}
\end{equation}
[which is a quantity not given explicitly in BS2, but follows from Eq.~(120) of that work]
and
\begin{align}
\wh \cD_L^\nd &= - \wh{\overline{\cM}}{}_{2,L}^\nd \wh G^\nd \wh{\overline{\cM}}{}_{2,L}^\nd
\frac1{1+\wh G^\nd  \wh{\overline{\cM}}{}_{2,L}^\nd }\,.
\label{eq:DL}
\end{align}

The output from these equations is a matrix, $\wh \cM_{3;\nd}$, in which the same underlying
quantity (the desired amplitude $\cM_{3;\nd}$) is expressed in terms of different variables,
as in Eq.~(\ref{eq:Kdf3tmat}). We need only one of the nine entries to reconstruct $\cM_{3;\nd}$.
We also note that the action of the symmetrization operators is perhaps better thought of
as ensuring that all diagrams are included in $\cM_{3;\nd}$. For example, if in
Eq.~(\ref{eq:symMdf3L}) we focus on the $\wh\cD_{23,L}^\nd \wh F^\nd$ term on the left end,
and keep only the leading term 
in the geometric series defining $\wh\cD_{23,L}^\nd$, 
then we are attaching a factor of  $-\wh{\overline{\cM}}{}_{2,L}^\nd\wh F^\nd$   
to $\wh{\wt{\cK}}_{\df,3;\nd}$.
Depending on which external flavor index we pick, we  get a contribution in which the
leftmost interaction involves scattering of a {\em different} pair, prior to the three-particle
(symmetric) interaction due to $\wt{\cK}_{\df,3;\nd}$. We must sum over the choices of interacting
pair in order to obtain the full expression for $\cM_3$, and this is implemented by
the symmetrization operators.

\section{$2d+1$ limit of nondegenerate formalism}
\label{sec:2d+1}

We consider in this section the limit of the nondegenerate formalism of BS2
(summarized in Appendix~\ref{sec:recap})
in which two of the three particles are
degenerate, $m_1=m_2$, and in which the interactions are invariant under
simultaneous $1\leftrightarrow 2$ exchange in the initial and final states.
We refer to this as a $2d+1$ system, with ``$d$'' standing for distinguishable,
and call the symmetry the ``partial flavor symmetry."
Examples of $2d+1$ systems are $\pi^+\pi^0 D_s^\pm$ in isosymmetric QCD.
These differ from the $2+1$ systems studied above,
such as $\pi^+\pi^+ D_s^\pm$,
because here the two degenerate particles are distinguishable, rather than identical.
Nevertheless, the two cases can be related using isospin symmetry.
In particular, the $\pi^+\pi^0$ subsystem can have isospin $1$ or $2$, corresponding
to antisymmetric or symmetric flavor wave functions.
The $I=2$ projection of the $2d+1$ quantization condition 
should be identical to that for the $2+1$ $\pi^+\pi^+ D_s^\pm$ system.
While this trick does not apply in all cases---for example, the $2+1$ $\pi^+\pi^+ K^+$ system
is related by isospin to a combination of $2d+1$ systems $\pi^+\pi^0 K^+$ and $\pi^+ \pi^+K^0$,
for which a multichannel formalism is needed---it does provide a check on the results
obtained in the main text.

Since this is a check, we only consider here the form of the quantization condition,
and associated relation of $\cK_{\df,3}$ to $\cM_3$, that
are most likely to be useful in practice, namely those involving the symmetrized three-particle K matrix.
These were derived in Sec.~\ref{sec:2+1symm} for the $2+1$ system.

\subsection{Quantization condition}

We begin by making a change compared to the nondegenerate analysis summarized in Appendix~\ref{sec:recap}. In the generic analysis following Eq.~(\ref{eq:momcons}) we chose to always
order the flavor labels $i$, $j$, and $k$ cyclically, a choice also made in the definition of
$\wt F^{(i)}$, Eq.~(\ref{eq:Ft}).
Here we make a different choice:
if the spectator flavor is $i=2$, then we take $j=1$ and $k=3$,
whereas for other spectator flavors the choices of $j$ and $k$ remain cyclic.
In words, if the spectator flavor is one of the degenerate pair, then the spherical harmonics
are always defined relative to the direction of the other member of the pair with this new choice.
This is the analog of the choice made in the analysis of $2+1$ systems, 
described in Sec.~\ref{subsec:onshell}.
It leads to the $1\leftrightarrow 2$ symmetry being fully manifest,
an example being that we now have the matrix identities
\begin{equation}
\wt F^{(2)} = \wt F^{(1)}\,, \quad
\wt G^{(12)}= \wt G^{(21)} = \wt G^{(12) \rm Tr}\,,\quad
\wt G^{(13)}= \wt G^{(23)}=\wt G^{(31) \rm Tr}=\wt G^{(32) \rm Tr}\,.
\label{eq:matrixidents}
\end{equation}
We stress that, while minor, this change ripples through the entire analysis, in ways that
we make explicit in the following.

With this convention, the $2d+1$ limits of $\wh F^\nd$ and $\wh G^\nd$ 
[defined in Eqs.~(\ref{eq:Fhat}) and (\ref{eq:Ghat}), respectively] are
\begin{align}
\wh F^\nd &\to {\rm diag}\left( \wt F^{(1)}, \wt F^{(1)}, \wt F^{(3)}\right)\,,
\label{eq:F2d+1}
\\
\wh G^\nd &\to \begin{pmatrix}
0 & \wt G^{(11)}  & P_L\wt G^{(13)} \\
\wt G^{(11)} & 0 & P_L \wt G^{(13)}P_L \\
\wt G^{(31)} P_L & P_L \wt G^{(31)} P_L & 0
\end{pmatrix}\,,
\label{eq:G2d+1}
\end{align}
where we have introduced $\wt G^{(11)} \equiv \wt G^{(12)}|_{m_1=m_2}$,
a quantity that has already appeared in Eq.~(\ref{eq:Ghat2p1}).
The change in flavor ordering has led to a rearrangement of the factors of $P_L$
compared to Eq.~(\ref{eq:Gt}).
The two-particle scattering matrices simplify similarly:
\begin{align}
\wh{\overline{\cK}}_{2,L,\nd} &\to {\rm diag}\left( 
\overline{\cK}_{2,L}^{(1)}, \overline{\cK}_{2,L}^{(1)}, \overline{\cK}_{2,L}^{(3)}\right)\,,
\\
\wh{\overline{\cM}}_{2,L,\nd} &\to {\rm diag}\left( 
\overline{\cM}_{2,L}^{(1)}, \overline{\cM}_{2,L}^{(1)}, \overline{\cM}_{2,L}^{(3)}\right)\,.
\end{align}

The effect of imposing the partial flavor symmetry on $\wh{\wt{\cK}}_{\df,3;\nd}$,
Eq.~(\ref{eq:Kdf3tmat}), is slightly more subtle.
Since $1\leftrightarrow 2$ interchange also changes the choice of particle that defines
the spherical harmonics, the symmetry becomes
\begin{equation}
\wh{\wt{\cK}}_{\df,3;\nd} 
= \matS \wh{\wt{\cK}}_{\df,3;\nd} \matS\,,
\qquad
\matS = \begin{pmatrix} 0 & 1 & 0 \\ 1 & 0 & 0 \\ 0 & 0 & P_L \end{pmatrix} = \matS^{\rm Tr}\,.
\label{eq:KdfS2}
\end{equation}
Implementing this, we find that the three-particle K matrix must have the form
\begin{equation}
\wh{\wt{\cK}}_{\df,3;\nd} =
 \begin{pmatrix} 
 \wt \cK_{\df,3;\nd}^{(11)} & \wt \cK_{\df,3;\nd}^{(12)} & \wt \cK_{\df,3;\nd}^{(13)} \\
 \wt \cK_{\df,3;\nd}^{(12)} & \wt \cK_{\df,3;\nd}^{(11)} & \wt \cK_{\df,3;\nd}^{(13)} P_L \\
 \wt \cK_{\df,3;\nd}^{(31)} & P_L \wt \cK_{\df,3;\nd}^{(31)} 
                & \wt \cK_{\df,3;\nd;ee}^{(33)} + \wt \cK_{\df,3;\nd;oo}^{(33)}
\end{pmatrix}\,,
\end{equation}
where the subscripts $e$ and $o$ refer, respectively, to
keeping only the even or odd angular-momentum entries.
We also know from its definition in Ref.~\cite{\BSnondegen}
that $\wh{\wt{\cK}}_{\df,3;\nd}$ is a real, symmetric matrix, which implies
that $\wt \cK_{\df,3;\nd}^{(11)}$, $\wt \cK_{\df,3;\nd}^{(12)}$,
and $\wt \cK_{\df,3;\nd}^{(33)}$ are symmetric (in the $\{k\ell m\}$ indices), 
and that $\wt \cK_{\df,3;\nd}^{(31)} = \wt \cK_{\df,3;\nd}^{(13){\rm Tr}}$.

To formalize the symmetry of the $2d+1$ system, we note that
in addition to $\wh{\wt{\cK}}_{\df,3,\nd}$,
the matrices $\wh F^\nd$, $\wh G^\nd$, and $\wh{\overline{\cK}}_{2,L,\nd}$
are invariant under conjugation by $\matS$, or, equivalently, commute with $\matS$.
To show this one needs the results that $P_L^2=1$,
$P_L  \overline{\cK}_{2,L,\nd}^{(j)} P_L= \overline{\cK}_{2,L,\nd}^{(j)}$
(which follows from angular momentum conservation),
and $P_L \wt F^{(3)} P_L = \wt F^{(3)}$ 
(which is a simple generalization of a result shown in Ref.~\cite{\HSQCa}).
Since $\matS^2=\mathbb 1$, the symmetry group is $S_2\cong Z_2$,
which has two irreps, denoted $\pm$, that are, respectively,
symmetric and antisymmetric under $1\leftrightarrow 2$ exchange.

The symmetric quantization condition for the $2d+1$ system is simply given by 
inserting the above-described forms for $\wh F^\nd$, $\wh G^\nd$, $\wh{\wt{\cM}}_{2,L,\nd}$ 
and $\wh{\wt{\cK}}_{\df,3;\nd}$ into Eq.~(\ref{eq:QC3ndsym}).
Since all elements of the quantization condition are invariant under $S_2$ transformations,
it can be block-diagonalized into parts transforming in the two irreps of $S_2$.
The corresponding projectors are:
\begin{equation}
\mathbb P_+ = \frac{\mathbb 1+ \matS}2 =
\ket{+}\bra{+} + \ket{3}\mathbb P_e \bra3
\quad
\ \ {\rm and}\ \
\quad
\mathbb P_- = \frac{\mathbb 1- \matS}2 =
\ket{-}\bra{-} + \ket{3}\mathbb P_o \bra3\,,
\end{equation}
where $\mathbb P_{e/o} = (1\pm P_L)/2$ projects onto even/odd angular momenta, and
\begin{equation}
\bra{\pm} = \left( \tfrac1{\sqrt2}, \pm \tfrac1{\sqrt2},0\right)
\ \ {\rm and}\ \
\bra3 = \left(0, 0, 1\right)\,.
\end{equation}
Since there are two independent projectors associated with each of $\mathbb P_\pm$,
the resulting quantization conditions for each irrep can be written in terms
of two-dimensional matrices. Both have the standard form
\begin{align}
0 &=\det\left(1+  \wh F_{3,\pm}
\wh{\wt{\cK}}_{\df,3,\pm}\right) \,,
\label{eq:QC3pm}
\\
\wh F_{3,\pm} &=
\left[
\frac{\wh F_\pm}3 - 
\wh F_\pm \frac1{\wh{\overline{\cK}}{}_{2,L,\pm}^{-1} + \wh F_\pm+ \wh G_\pm} \wh F_\pm
\right]\,,
\label{eq:F3hatpm}
\end{align}
with the component matrices given by
\begin{equation}
\begin{gathered}
\wh F_\pm = {\rm diag} \left( \wt F^{(1)}, \wt F^{(3)}_{ee/oo}\right)\,,\quad
\wh G_\pm = \begin{pmatrix}
\pm \wt G^{(11)} & \sqrt2 P_L \wt G^{(13)} \mathbb P_{e/o} \\
\sqrt2  \mathbb P_{e/o} \wt G^{(31)} P_L & 0
\end{pmatrix}\,,
\label{eq:newFG}
\\
\wh{\overline{\cK}}_{2,L,\pm} = 
{\rm diag} \left( \overline \cK_{2,L}^{(1)}, \overline \cK^{(3)}_{2,L,ee/oo}\right)\,,\quad
\wh{\wt{\cK}}_{\df,3,\pm} = \begin{pmatrix}
\wt \cK_{\df,3;\nd}^{(11)} \pm \wt \cK_{\df,3;\nd}^{(12)} 
& \sqrt{2} \wt \cK_{\df,3;\nd}^{(13)} \mathbb P_{e/o} \\
\sqrt{2} \mathbb P_{e/o} \wt \cK_{\df,3;\nd}^{(31)} & \wt \cK_{\df,3;\nd;ee/oo}^{(33)} 
\end{pmatrix} \,.
\end{gathered}
\end{equation}
We observe two differences between the results for the two irreps: the sign of the $\wt G^{(11)}$
contribution, and the nature of the angular-momentum projection for the third index 
(onto even or odd angular momenta).

We now argue that the quantization condition for the symmetric irrep, 
i.e.~Eq.~(\ref{eq:QC3pm}) composed of quantities with subscripts $+$,
is identical to that for the $2+1$ system given in Eq.~(\ref{eq:QC3pipiK}).
To do so, we need to show that 
\begin{equation}
\wh F_+=\wh F\,, \quad
\wh G_+=\wh G\,, \quad
\wh{\overline \cK}_{2,L,+} = \wh{\overline \cK}_{2,L}\,, \ \ {\rm and}\ \
\wh{\wt \cK}_{\df,3,+}=\wh \cK_{\df,3}\,.
\label{eq:equalities}
\end{equation}
where the $2+1$ quantities on the right-hand sides of these equalities are
defined in Eqs.~(\ref{eq:Fhat2p1}), (\ref{eq:Ghat2p1}),
(\ref{eq:K2Lhat2p1}), and Eq.~(\ref{eq:Kdf3hat2p1}), respectively.
Here we focus on the first three of these equalities; the final one is demonstrated in the
next subsection.

First we consider the desired equalities $\wh F_+=\wh F$  and $\wh G_+=\wh G$.
There is a notational difference in that here we label the third flavor as $3$, whereas
in the $2+1$ analysis it is called $2$, but this is a trivial difference as the label is a dummy
variable.
The only nontrivial difference is the 
presence of projectors $\mathbb P_e$ acting on the second index in $\wh F_+$ and
$\wh G_+$.
However, this projection could equally well be included also
in $\wh F$ and $\wh G$, because the quantities that these matrices 
multiply---two- and three-particle K matrices for the $2+1$ system---are symmetric
under interchange of the identical particles and thus contain only even partial waves.

Turning to the desired equality $\wh{\overline \cK}_{2,L,+} = \wh{\overline \cK}_{2,L}$,
we observe an apparent difference due to the factor of $1/2$ in the second
entry of latter quantity [see Eq.~(\ref{eq:K2Lhat2p1})]. 
However, this is canceled by a factor of $2$ in the relation
\begin{equation}
\overline{\cK}_{2,L}^{(2)} = 2 \overline{\cK}_{2,L,ee}^{(3)}\,,
\label{eq:K2Lequality}
\end{equation}
which we now explain. 
The quantity on the left-hand side contains, 
aside from kinematic factors common to both sides
(because, as above, flavor 2 on the left-hand side is the same as flavor 3 on the right),
the two-particle K matrix for identical particles (e.g.~$\pi^+\pi^+$ scattering). 
This contains only even partial waves, matching the $ee$ projection on the right-hand side.
Furthermore, it is equal to the amplitude for the symmetric projection of the two degenerate
but distinguishable particles (e.g.~$(\pi^+\pi^0)_{I=2}$) as long as one uses normalized
in and out states, created by operators of the form 
$(\pi^+(\bm p_1) \pi^0(\bm p_2) +\pi^+(\bm p_2) \pi^0(\bm p_1))/\sqrt2$.
When evaluating the amplitude between two such operators, there are four contractions,
for each of which the even partial waves are given by $\overline{\cK}_{2,L,ee}^{(3)}$.
Combining with the $(1/\sqrt 2)^2$ one finds an overall factor of $2$.

Finally, we note that the three equalities that we have established show that
\begin{equation}
 \wh F_{3,+} = \wh F_3\,,
\label{eq:F3equality}
\end{equation}
where $\wh F_3$ is defined in Eq.~(\ref{eq:F3hat2p1}).

\subsection{Relation of $\cK_{\df,3}$ to $\cM_3$}

In this subsection, we demonstrate the equality $\wh \cK_{\df,3} = \wh{\wt{\cK}}_{\df,3,+}$
by relating both K matrices to  the same physical scattering amplitude.
This completes the demonstration of the equivalence of the
quantization condition in Eq.~(\ref{eq:QC3pm}) to the $2+1$ version, Eq.~(\ref{eq:QC3pipiK}).

The result we need in the $2+1$ theory is given by Eqs.~(\ref{eq:M3Ldecomp}) and
(\ref{eq:final1}), which express $\cM_{3}$ in that theory in terms of $\wh \cK_{\df,3}$.
[This requires taking the standard $L\to\infty$ limit of Eq.~(\ref{eq:M3Ldecomp}).]
In the $2d+1$ theory, we must again decompose into irreps of $S_2$, 
but now doing so for the amplitude $\cM_{3;\nd}$.
In the $\pi^+\pi^0 D_s^\pm$ example this corresponds to picking out the
amplitudes in which the two pions have $I=2$ or $I=1$.
Since the $2+1$ amplitude in this example, $\pi^+\pi^+ D_s^\pm$, is pure $I=2$,
in isosymmetric QCD we have the relation
$\cM_3 = \cM_{3;\nd}^{I=2}$.
The challenge is thus to express $\cM_{3;\nd}^{I=2}$ in terms of the appropriate
projection of $\cM_{3;\nd}$.

To do so, it is useful to rewrite the expression for the matrix $\wh \cM_{3,L;\nd}$,
Eq.~(\ref{eq:M3hatresb}), as a result for the amplitude itself, 
since each element of the matrix contains the same underlying quantity. 
This is achieved by
\begin{equation}
\cM_{3,L;\nd}(\{\bm p\};\{\bm k\}) = \bra{1_\cS} \left(
\wh \cM'_{\df,3,L;\nd} + \wh \cD_L\right) \ket{1_\cS}\,,
\label{eq:M3hatresn}
\end{equation}
where $\bra{1_\cS}$ and $\ket{1_\cS}$ are similar to the vectors 
$\bra1=(1,1,1)$ and $\ket{1}=(1,1,1)^{\rm Tr}$,
except that they implicitly include the operation of combining with appropriate spherical harmonics 
before taking the inner products.
We also stress that, throughout this subsection, we continue to work in the $2d+1$ limit of the
nondegenerate theory, although the subscripts are kept as ``nd'' alone, in order to lighten the
notation.

We next introduce $\mathbb 1 =\mathbb P_++\mathbb P_-$,
which block-diagonalizes the matrices on the right-hand side, 
\begin{align}
\cM_{3,L;\nd} &= \cM_{3,L,+} + \cM_{3,L,-} \,,
\label{eq:M3hatresnn}
\\
\cM_{3,L,\pm} &=  
\bra{1_\cS} \mathbb P_\pm\left(
\wh \cM'_{\df,3,L;\nd} + \wh \cD_L \right) \mathbb P_\pm \ket{1_\cS}\,.
\label{eq:M3hatresnnn}
\end{align}
We have dropped the momentum arguments on the left-hand side for brevity.
Inserting the explicit forms for $\mathbb P_\pm$ from above, 
the projected amplitudes can be rewritten in terms
of $2\times2$ matrices that act in the same space as in the previous subsection:	
\begin{align}
\cM_{3,L,\pm} &= \left(\braket{1_\cS}{\pm}, \braket{1_\cS}{3} \mathbb P_{e/o}\right)
\wh \cM_{3,L,\pm}
\begin{pmatrix} \braket{\pm}{1_\cS} \\ \mathbb P_{e/o} \braket{3}{1_\cS}
\end{pmatrix} \,,
\label{eq:M3Lpm}
\\
\wh \cM_{3,L,\pm} &=
\begin{pmatrix}
\bra{\pm}\wh \cM''_{\df,3,L} + \wh \cD_L\ket{\pm} &
\bra{\pm}\wh \cM''_{\df,3,L} + \wh \cD_L\ket{3}
\\
\bra{3}\wh \cM''_{\df,3,L} + \wh \cD_L\ket{\pm} &
\bra{3}\wh \cM''_{\df,3,L} + \wh \cD_L\ket{3}
\end{pmatrix}
\\
&=
\wh \cM''_{\df,3,L,\pm} + \wh \cD_{L,\pm}\,,
\end{align}
where it follows from Eqs.~(\ref{eq:symMdf3L}), (\ref{eq:DL}), (\ref{eq:D23L}) and (\ref{eq:M2Lhat}) that
\begin{align}
\wh \cM''_{\df,3,L,\pm} &=
\left[\tfrac13 - \wh \cD_{23,L,\pm}  \wh F_\pm\right] 
\wh{\wt{\cK}}_{\df,3,\pm} 
\frac1{1+ \wh F_{3,\pm} \wh{\wt{\cK}}_{\df,3,\pm}}
\left[\tfrac13 - \wh F_{\pm}  \wh \cD_{23,L,\pm}\right]\,,
\label{eq:Mdf3Lpm}
\\
 \wh \cD_{L,\pm} &=
- \wh{\overline{\cM}}_{2,L,\pm} \wh G_\pm \wh{\overline{\cM}}_{2,L,\pm}
\frac1{1+\wh G_\pm  \wh{\overline{\cM}}_{2,L,\pm}}\,,
\\
\wh \cD_{23,L,\pm} &=  \wh{\overline{\cM}}_{2,L,\pm}
\frac1{1+\wh G_\pm \wh{\overline{\cM}}_{2,L,\pm} }\,,
\\
\wh{\overline{\cM}}_{2,L,\pm}^{-1} &= \wh{\overline{\cK}}_{2,L,\pm}^{-1} + \wh F_\pm\,.
\end{align}

We now observe that
\begin{equation}
\left(\braket{1_\cS}{+}, \braket{1_\cS}{3} \mathbb P_{e}\right)
 =
\frac{\bra{\alpha_\cS}}{\sqrt2}\,,
\qquad
\begin{pmatrix} \braket{+}{1_\cS} \\ \mathbb P_{e} \braket{3}{1_\cS} \end{pmatrix}
 =
\frac{\ket{\alpha_\cS}}{\sqrt2}\,,
\label{eq:useful}
\end{equation}
where $\bra{\alpha_\cS}$ and $\ket{\alpha_\cS}$ are defined in the discussion following
Eq.~(\ref{eq:M3L2p1final}). 
The first result follows because the symmetrization operator
$\cS_{11'}$ contained in $\bra{\alpha_\cS}$ is represented 
by the inner product with $\sqrt 2\bra+$
on the upper two coordinates in the $3\times3$ matrix form, and by $2\mathbb P_e$
on the third coordinate.
A similar argument holds for the second result in Eq.~(\ref{eq:useful}).
Thus we obtain 
\begin{equation}
\cM_{3,L,+} = \frac12\bra{\alpha_\cS} \wh \cM_{3,L,+} \ket{\alpha_\cS}\,,
\label{eq:M3Lpfinal}
\end{equation}
which, when, $L\to\infty$, yields an equation for $\cM_{3,+}$, the symmetric part of $\cM_{3;\nd}$.

From the discussion at the start of this subsection, 
we know that the $2+1$ theory three-particle amplitude, $\cM_3$,
is equal to $\cM_{3,\nd}^{I=2}$. The latter is, in turn,
proportional to $\cM_{3,+}$. To determine
the proportionality constant we argue as in the demonstration of Eq.~(\ref{eq:K2Lequality}) above.
To obtain the $[\pi^+\pi^0]_{I=2} D_s^+$ amplitude, 
we must use external two-pion operators of the form
$(\pi^+(\bm p_1) \pi^0(\bm p_2) +\pi^+(\bm p_2) \pi^0(\bm p_1))/\sqrt2$.
This combination picks out the symmetric part of $\cM_{3;\nd}$,
once for each of the four contractions, with an overall factor of $(1/\sqrt2)^2$ from the operators.
Thus the proportionality constant is $2$,
\begin{equation}
\cM_3 = \cM_{3,\nd}^{I=2} = 2 \cM_{3,+}\,.
\end{equation}
Using Eq.~(\ref{eq:M3Lpfinal}) we thus obtain
\begin{equation}
\cM_3 = 2 \lim_{L\to\infty} \cM_{3,L,+} =
\lim_{L\to\infty} \bra{\alpha_\cS} \wh \cM_{3,L,+} \ket{\alpha_\cS}\,.
\label{eq:M3toM3p}
\end{equation}
This result relates the $2+1$ amplitude on the left-hand side to a $2d+1$ expression on the right.

We now have all the ingredients to obtain the desired equality of three-particle K matrices.
We must simply equate the expression for $\cM_3$ 
that follows from Sec.~\ref{subsec:QCsymm2p1}, namely
\begin{align}
\cM_3 &= \lim_{L\to\infty} \bra{\alpha_\cS} 
\left(\wh \cM_{\df,3,L}^{\uu\prime\prime} + \wh D_L^\uu  \right)\ket{\alpha_\cS}\,,
\end{align}
with that for the right-hand side of Eq.~(\ref{eq:M3toM3p}), which, from above, is
\begin{equation}
\lim_{L\to\infty} \bra{\alpha_\cS} 
\left(\wh \cM_{\df,3,L,+} + \wh \cD_{L,+}\right)\ket{\alpha_\cS}\,.
\end{equation}
The second terms in both expressions are equal, because
the equalities in Eq.~(\ref{eq:equalities}) imply that
\begin{equation}
\wh{\overline \cM}_{2,L,+} = \wh{\overline \cM}_{2,L}\,,\quad
\wh \cD_{23,L,+} = \wh \cD_{23,L}^\uu\,, \ \ {\rm and} \ \
\wh \cD_{L,+} = \wh \cD_L^\uu\,,
\end{equation}
where the $2+1$ theory quantities on the right-hand sides are given in
Eqs.~(\ref{eq:M2Lhat2p1}),
(\ref{eq:D23Lhat}),
and
(\ref{eq:DLhat}),
respectively.
Furthermore, the expressions for $\wh \cM_{\df,3,L}^{\uu\prime\prime}$
[Eq.~(\ref{eq:final1})]
and $\wh \cM_{\df,3,L,+}$
[Eq.~(\ref{eq:Mdf3Lpm})]
agree term by term, but only if $\wh \cK_{\df,3} = \wh{\wt {\cK}}_{\df,3,+}$.
This establishes the desired result.

For completeness, we give an expression for the antisymmetric part of the amplitude
that is analogous to Eq.~(\ref{eq:M3Lpfinal}).
To do so we need to introduce
antisymmetrization operators $\overrightarrow{\cA}_{12}$ and $\overleftarrow{\cA}_{12}$,
in terms of which\footnote{%
One might naively think that $\braket{1_\cS}{-}=0$, since $\bra{1_\cS}$ implies summing over
all entries, while $\ket-$ is antisymmetric. This is incorrect because the components are
not being subtracted as matrices with $\{k\ell m\}$  indices, but rather are to be first combined
with spherical harmonics and then subtracted. This simply leads to the desired antisymmetric
projection.}
\begin{equation}
\left(\braket{1_\cS}{-}, \braket{1_\cS}{3} \mathbb P_{o}\right)
 =
 \left(\overrightarrow{\cA}_{12} /\sqrt2, \overrightarrow{\cA}_{12}/2\right)
\equiv
\frac{\bra{\alpha_\cA}}{\sqrt2}\,,
\end{equation}
and similarly for the corresponding ket.
Thus we obtain
\begin{equation}
\cM_{3,L,-} = \frac12\bra{\alpha_\cA} \wh \cM_{3,L,-} \ket{\alpha_\cA}\,,
\label{eq:M3Lmfinal}
\end{equation}
which, when $L\to\infty$, leads to the usual type of integral equations.

\bibliography{ref} 

\end{document}